\documentclass[twocolumn]{aastex631}

\usepackage{mathptmx}
\usepackage[T1]{fontenc}
\usepackage{ae,aecompl}
\usepackage{graphicx}	
\graphicspath{{./Figures/}} 
\usepackage{amsmath}	
\usepackage{amssymb}	
\usepackage{subfigmat}
\usepackage{enumerate}
\usepackage{multirow}
\usepackage{natbib}

\newcommand{\kms}{\hbox{km s$^{-1}$}}

\newcommand{\chxoh}{\mbox{CH$_3$OH}}
\newcommand{\xchxoh}{\mbox{$^{13}$CH$_3$OH}}
\newcommand{\funit}{\mbox{mJy~beam$^{-1}$}}
\newcommand{\funitv}{\mbox{mJy~beam$^{-1}$~km~s$^{-1}$}}

\usepackage{tabularx, booktabs}
\usepackage{rotating}
\usepackage{longtable}
\usepackage{ltablex}
\usepackage{threeparttablex}

\def\gtrsim{\mathrel{\hbox{\rlap{\hbox{\lower5pt\hbox{$\sim$}}}\hbox{$>$}}}}
\newcommand{\msun}{\mbox{M$_\odot$}}
\newcommand{\lsun}{\mbox{L$_\odot$}}

\newcommand{\degree}{\mbox{$^\circ$}}

\newcolumntype{Y}{>{\centering\arraybackslash}X}

\shorttitle{Complex Organic Molecules in HOPS 373SW}
\shortauthors{Lee et al.}

\begin{document}

\title{Complex Organic Molecules in a Very Young Hot Corino, HOPS 373SW}

\correspondingauthor{Jeong-Eun Lee}
\email{lee.jeongeun@snu.ac.kr}

\author[0000-0003-3119-2087]{Jeong-Eun Lee}
\affiliation{Department of Physics and Astronomy, Seoul National University, 1 Gwanak-ro, Gwanak-gu, Seoul 08826, Korea}

\author{Giseon Baek}
\affiliation{Department of Physics and Astronomy, Seoul National University, 1 Gwanak-ro, Gwanak-gu, Seoul 08826, Korea}

\author{Seokho Lee}
\affiliation{Korea Astronomy and Space Science Institute, 776 Daedeok-daero, Yuseong, Daejeon 34055, Korea}

\author{Jae-Hong Jeong}
\affil{Department of Physics and Astronomy, Seoul National University, 1 Gwanak-ro, Gwanak-gu, Seoul 08826, Korea}

\author{Chul-Hwan Kim}
\affiliation{Department of Physics and Astronomy, Seoul National University, 1 Gwanak-ro, Gwanak-gu, Seoul 08826, Korea}

\author{Yuri Aikawa}
\affiliation{Department of Astronomy, University of Tokyo, 7-3-1 Hongo, Bunkyo-ku, Tokyo 113-0033, Japan}

\author{Gregory J. Herczeg}
\affiliation{Kavli Institute for Astronomy and Astrophysics, Peking University, Yiheyuan 5, Haidian Qu, 100871 Beijing, China}
\affiliation{Department of Astronomy, Peking University, Yiheyuan 5, Haidian Qu, 100871 Beijing, China}

\author[0000-0002-6773-459X]{Doug Johnstone}
\affiliation{NRC Herzberg Astronomy and Astrophysics, 5071 West Saanich Rd, Victoria, BC, V9E 2E7, Canada}
\affiliation{Department of Physics and Astronomy, University of Victoria, Victoria, BC, V8P 5C2, Canada}

\author{John J. Tobin}
\affiliation{National Radio Astronomy Observatory, 520 Edgemont Rd., Charlottesville, VA 22903}

\begin{abstract}
We present the spectra of Complex Organic Molecules (COMs) detected in HOPS 373SW with the  Atacama  Large  Millimeter/submillimeter  Array (ALMA). HOPS 373SW, which is a component of a protostellar binary with a separation of 1500 au, has been discovered as a variable protostar by the JCMT~Transient monitoring survey with a modest ($\sim30\%$) brightness increase at submillimeter wavelengths. 
Our ALMA Target of Opportunity (ToO) observation at $\sim$345 GHz for HOPS 373SW revealed extremely young chemical characteristics with strong deuteration of methanol.
The dust continuum opacity is very high toward the source center, obscuring line emission from within 0.03\arcsec.
The other binary component, HOPS 373NE was detected only in C$^{17}$O in our observation, implying a cold and quiescent environment.
We compare the COMs abundances relative to \chxoh in HOPS 373SW with those of V883 Ori, which is an eruptive disk object, as well as other hot corinos, to demonstrate the chemical evolution from envelope to disk. High abundances of singly, doubly, and triply deuterated methanol (CH$_2$DOH, CHD$_2$OH, and CD$_3$OH) and a low CH$_3$CN abundance
in HOPS 373SW compared to other hot corinos suggest a very early evolutionary stage of HOPS 373SW in the hot corino phase.
Since the COMs detected in HOPS 373SW would have been sublimated very recently from grain surfaces, HOPS 373SW is a promising place to study the surface chemistry of COMs in the cold prestellar phase, before sublimation.

\end{abstract}


\section{Introduction}

Comets and asteroids reveal that the Solar nebula was rich in water and organic molecules. The recent Rosetta mission also showed that many complex organic molecules (COMs, carbon-bearing molecules made of 6 or more atoms), as well as so-called prebiotic molecules, which could be precursors of amino acids and sugars, exist in the comet 67P/C-G \citep{Altwegg17}.
These organics, together with water, could have been brought to the young Earth's surface by comets and asteroids, which are believed to preserve some compositions from the early stage of star formation. Therefore, the investigation of COMs in the circumstellar environments of Sun-like young stellar objects is important to infer better the chemical distribution in our Solar system.

Significant progress has been made recently thanks to the ALMA observations of low-mass protostars in various transitions of COMs \citep{Jorgensen2016, Imai16, Lopez17, Sahu19, Jacobsen19, Yang20, Belloche20, vanGelder20, Yang21,Hsu22}.
The hot corinos, where forming stars heat the surrounding material above the sublimation temperature (100 K) of the water ice mantles of dust grains, are very rich in COMs  \citep{Ceccarelli07},
although not all protostars are observed to develop hot corinos.  
Hot corinos are discovered predominantly toward the youngest and most embedded Class 0 protostars with bolometric luminosities greater than 5 \lsun, mostly \citep{Belloche20,Yang20,Hsu22}. On the other hand, COMs as rich as those in hot corinos have been rarely detected in disks, except for special cases such as the disk surface of HH 212 \citep{cflee19}
and the eruptive disk midplane of V883 Ori \citep{van'tHoff18,jelee19}.

Chemical models for COMs require both gas-phase and ice-phase processes in protostellar environments, although ice chemistry is essential to the formation of COMs.  
During the warm-up phase (20--40 K) developed by protostellar accretion, COMs form efficiently via diffusion of radicals and molecules on grains \citep{Garrod06}. However, in cold prestellar cores,  other mechanisms, such as cosmic-ray-induced chemistry \citep{Shingledecker18} and non-diffusive reactions \citep{Chang16,mhJin20}, may be important to explain the observed abundances of COMs. We need to explore how organic complexity evolves over the protostellar lifetime to resolve the details of how the formation chemistry of COMs proceeds.
Therefore, observations and comparisons of COMs across a wide range of evolutionary stages are critical.

Recently, the ALMA Survey of Orion Planck Galactic Cold Clumps (ALMASOP) project for 11 Class 0 hot corinos clearly showed that the bolometric luminosity positively correlated with the extent and the total number of gas-phase methanol, supporting the thermal sublimation of COMs in the hot corinos \citep{Hsu22}. In addition, the abundances of specific COMs with respect to \chxoh\ in these hot corinos are homogeneous, inferring that chemistry does not evolve much over the short timescale of evolution within Class 0 protostars.

According to \citet{jelee19}, the COMs in the disk midplane of V883 Ori do show some hints of chemical evolution with higher abundances compared to the hot corino IRAS 16293-2422B (hereafter IRAS 16293B). V883 Ori is in transition from Class I to Class II \citep{jelee19}. Therefore, comparisons of relative abundances of COMs over a longer evolutionary timescale, covering the very beginning stage of Class 0 through Class II, are feasible to explore the chemical evolution of COMs.
Unfortunately, detecting COMs in both extremely young Class 0 objects with very low luminosities as well as Class II objects is challenging because their luminosities are not high enough to develop a detectable region of COMs emission. 
One desirable solution is to utilize accretion bursts, which can extend the COMs sublimation region to larger radii, as demonstrated in V883 Ori.


Episodic accretion \citep{hartmann96} is now accepted as the standard accretion model in the low mass star formation, and thus, the burst accretion is a common phenomenon throughout the protostellar evolution with more frequent and more powerful accretion bursts in earlier evolutionary stages.
Recently, surveys at long wavelengths \citep[e.g.][]{scholz13,rebull14,antoniucci14,lucas17,Johnstone2018,fischer19,lee21, park21,zakri22} have evaluated the statistics of accretion variability during the earlier stages of protostellar evolution. Therefore, we are now in a better situation to study COMs even in the earliest 
protostars.

HOPS 373, a young protostellar binary system with a separation of 1500 AU \citep{tobin15,tobin20}, was discovered as a variable with $\sim30\%$ brightness increase at 850 $\mu$m by the JCMT~Transient monitoring survey \citep{yoon2022}. HOPS 373 has been classified as a PACS Bright Red Source \citep{Stutz2013}, which is consistent with an early stage of a Class 0 protostar.  
Following this JCMT Transient discovery, we triggered our ALMA Target of Opportunity (ToO) program to study the sublimated COMs by the recent (2021 March) burst event  \citep[][]{yoon2022} in HOPS 373 with almost the same spectral setup used for V883 Ori \citep{jelee19}. The identical observations of two eruptive sources in very distinct evolutionary stages within the same star-forming environment of the Orion cloud provide us an important opportunity to study the chemical evolution of COMs from the protostellar envelope to the disk. 

\section{Observations}
 HOPS 373 was observed with ALMA during Cycle 7 (2019.1.00386T, PI: Jeong-Eun Lee) on 2021 August 2 and 6.
 The coordinates of the phase center are $\alpha=05^h46^m30^s.905$ and $\delta=-00^d02^m35^s.200$ (J2000).
Two observations were carried out with slightly different configurations: the earlier observation has the baseline range from 46.8 m to 5.6 km with thirty-eight 12-m antennas, while the later observation consists of the baselines from 70 m to 6.9 km with thirty-nine antennas. The total observing time on HOPS 373 was 30.5 minutes for each observation. 
 The eight spectral windows were set for 335.327-335.445 GHz, 336.485--336.719 GHz, 336.993--337.110 GHz, 338.306--338.775 GHz, 346.937--347.406 GHz, 349.162--349.396 GHz, 350.035--350.152 GHz, 350.353-- 350.470 GHz, each with a spectral resolution of 244.141 kHz ($\sim$0.2 \kms).
 This setup contains HDO $3_{3,1}-4_{2,2}$, C$^{17}$O 3--2, CH$_3$OH 7$_{\rm k}-6_{\rm k}$, H$^{13}$CO$^+$ 4--3, CCH N=~4--3, J=~7/2--5/2, F=~4--3 and F=~3--2 lines, and many COMs including isotopologues.

\begin{figure*}[htp]
\centering
\includegraphics[trim=0cm 0.5cm 0cm 0cm, clip=true,width=0.95\textwidth]{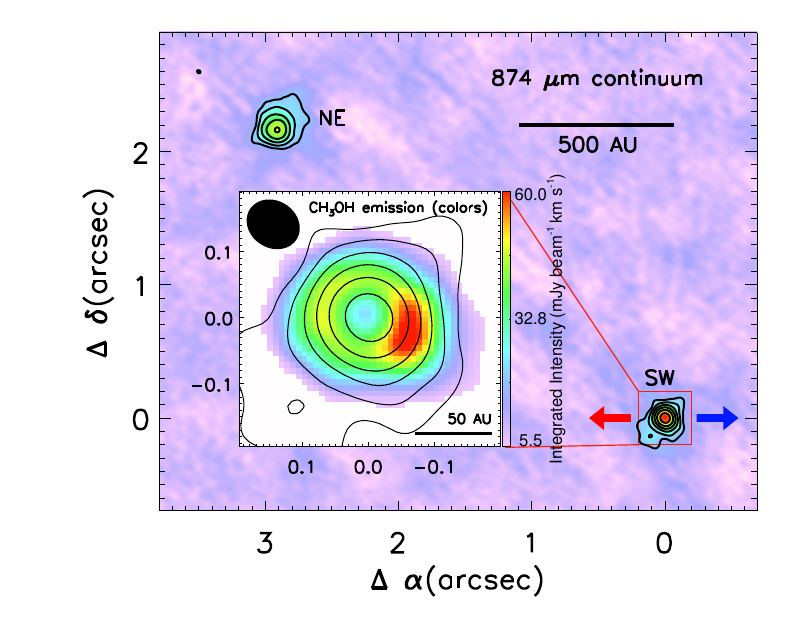}
\caption{The ALMA Band 7 continuum image. The contours are 10, 20, 40, 80, 160, and 320 $\sigma$ (1 $\sigma$ = 0.11 \funit). The red and blue arrows denote the outflow directions of red- and blue-shifted components, respectively, in a small scale in the vicinity of HOPS 373SW. The inset presents the integrated intensity map of the averaged \chxoh\, (color image) with the same contours for the continuum intensity. The \chxoh\, intensity is integrated within $\pm$1 \kms\, after correcting for the envelope velocity profile using the \xchxoh\ intensity weighted velocity map (the second top left panel of Figure~\ref{fig:coms_m0m1}), and only the integrated intensity over the 5 $\sigma$ (1 $\sigma$ = 1.1 \funitv) is presented in the color image. The observed beam is presented in the filled black ellipse on the top left corner.}
\label{fig:cont_mol}
\end{figure*}
 
 The data were initially calibrated using the CASA 6.1.1.15 pipeline \citep{McMullin2007}.
 The quasar J0510+1800 was used as a bandpass and amplitude calibrator,
and the nearby quasar J0532+0732 was used as a phase calibrator.
Only a phase self-calibration with a solution interval of `inf' was applied for better imaging.
Only the baselines with the UV distance longer than 100 m were used {\bf to remove} stripe features shown in the continuum image.

The natural weighting was used to obtain a high signal-to-noise (S/N) ratio as well as a high spatial resolution; the beam has a size of 0.08\arcsec\,$\times$ 0.07\arcsec\ (34 au $\times$ 30 au) with a  position angle of  $\sim$ 55\degree.
A dust continuum image was produced  using line-free channels. The RMS noise levels of the line cube and the continuum image are $\sim$3 and 0.11 \funit, respectively.\\

\begin{figure*}[htp]
\centering
\includegraphics[width=1.0\textwidth]{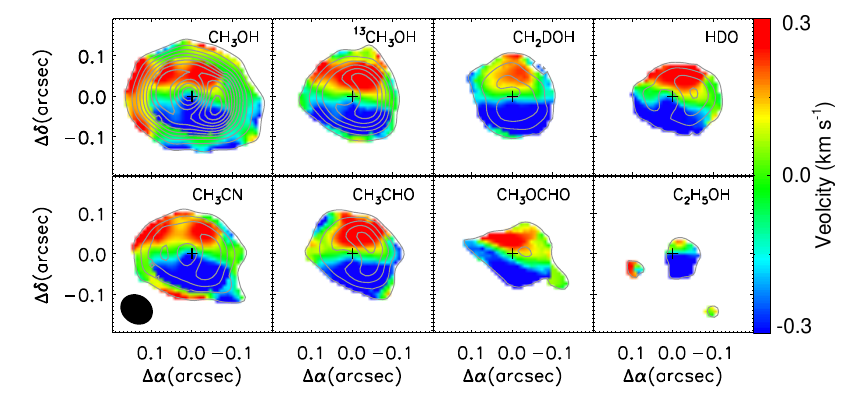}
\caption{Intensity weighted velocity maps (colors) and integrated intensity maps (contours) of 8 different molecular lines. Two isolated lines are averaged each for \chxoh, \xchxoh, and CH$_3$CHO. The start and step of contours are 5 $\sigma$. 
The rms noise levels are 1.1, 1.2, 2.1, 2.0, 2.3, 1.6, 1.9, and 2.3 \funitv\, for \chxoh, \xchxoh, CH$_2$DOH, HDO, CH$_3$CN, CH$_3$CHO, CH$_3$OCHO, and C$_2$H$_5$OH, respectively. 
The cross denotes the position of the continuum peak. The observed beam is presented in the lower left panel. The source velocity of 10.13 \kms\, is adopted}
\label{fig:coms_m0m1}
\end{figure*}

\section{Emission Distribution} \label{sec:emision_distribution}

\subsection{Continuum emission}

Both continuum sources in the binary system \citep{yoon2022} are well resolved, as presented in Figure \ref{fig:cont_mol}. In this observation, the flux densities of the NE and SW sources are 75.1 ($\pm1.4$) and 90.4 ($\pm1.4$) mJy at 0.87 mm, respectively. However, their previous flux densities were 85.0 ($\pm3.3$) and 81.8 ($\pm1.7$) mJy at 0.89 mm, respectively, in \citet{yoon2022}, where the two observations in 2016 and 2017 were averaged to increase the S/N ratio since the fluxes did not change much between the two observations.
Each flux density in two different observations cannot be directly compared because of two different observational parameters and the absolute calibration error \citep[see, e.g.,][]{Francis2020}. 
However, the flux ratio between two sources (SW/NE) within the same observation must be robust, and the ratio has increased from 0.96 to 1.20 by 
a factor of 1.25 ($\pm0.1$) 
supporting the SW source as the primary brightening source. 

The mass of each target may be estimated from the continuum flux by 
\begin{equation} \label{eq:one}
   M = \frac{F_\nu d^2}{\kappa_\nu B_\nu (T_{\rm dust})},
\end{equation}
where $F_\nu$ is the observed flux, $d$ is the distance to HOPS 373, 
$\kappa_\nu$ is the dust opacity, and $B_\nu (T_{\rm dust})$ is the Planck function 
with the dust temperature of $T_{\rm dust}$.  We adopt an average dust
temperature of 65 K by using the method of \citet{tobin20} and the bolometric luminosity of 5.3 $L_\odot$ \citep{kang15}, the OH5 dust opacity 
$\kappa_\nu =$ 1.84 cm$^{2}$~g$^{-1}$ \citep{Ossenkopf94}, and a gas-to-dust ratio of 100.
The estimated mass could be highly uncertain due to the continuum optical depth and the dust opacity \citep{Williams2011}.
The properties of two continuum sources are summarized in Table \ref{tb:gauss_cont}.

\subsection{Molecular emission}

Except for C$^{17}$O, all line emission is detected only in HOPS 373SW. 
The line emission is also affected significantly by the dust continuum opacity \citep{harsono18, jelee19}; as seen in the inset of Figure~\ref{fig:cont_mol}, the molecular emission of HOPS 373SW shows a hole inside the radius of $\sim$0.03\arcsec. Figure~\ref{fig:coms_m0m1} shows 8 molecular line intensity distributions (contours) on top of the intensity-weighted velocity maps (colors) in HOPS 373SW. All molecular emission decreases toward the center (due to the dust continuum opacity) with molecular emission peaks located $\sim$0.06\arcsec\ away from the center to the west. 
The blue-shifted outflow cavity is located to the west. Therefore, the western part of the emission faces less extinction. On the other hand, the emission in the east part is extinct by the thick envelope material beyond the red-shifted cavity wall.
The C$^{17}$O J=3--2 line is not present in Figure~\ref{fig:coms_m0m1} because it shows the inverse P-Cygni profile with the red-shifted absorption against the continuum, indicative of infall \citep{evans15}.

The intensity-weighted velocity maps of all molecular lines (Figure \ref{fig:coms_m0m1}) clearly show the velocity gradient from the north to the south, indicative of a rotational motion perpendicular to the known CO outflow direction \citep{yoon2022}. On the other hand, the \chxoh\, map (the upper left of Figure \ref{fig:coms_m0m1}) shows another velocity gradient consistent with the EW outflow direction beyond 0.1\arcsec\ (S. Lee et al.~submitted.).
Therefore, we extract the spectra from the emission area between 0.03\arcsec\ and 0.1\arcsec\ in the western semicircle to increase the S/N and minimize the outflow effect. Inside 0.03\arcsec, the dust continuum opacity is high enough to produce sub-continuum absorption in molecular lines. 

\begin{figure*}[htp]
\centering
\includegraphics[width=0.495\textwidth]{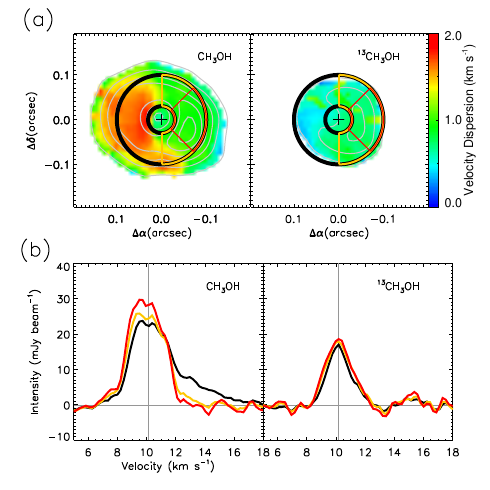}
\includegraphics[width=0.495\textwidth]{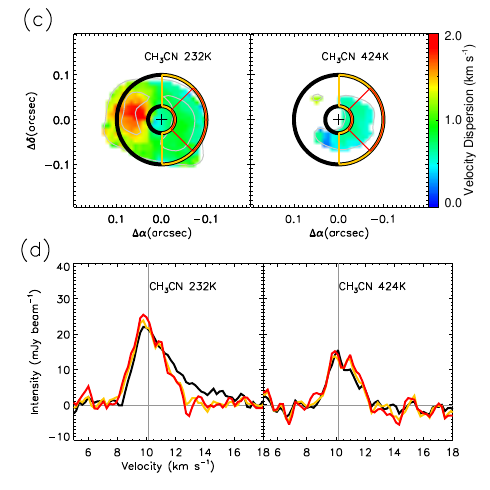}
\vspace{-0.8cm}
\caption{Velocity dispersion (an image) and the averaged spectra for the \chxoh\ and \xchxoh\ lines.(a) The contours indicate the integrated intensity and the start and step of contours are 10 $\sigma$. (b) The spectra are averaged within the regions outlined with the same colors (black, orange, and red) in the top panels. The vertical lines indicate the source velocity (10.13 \kms) measured from the \xchxoh\, spectra shown in the bottom right panel. (c) and (d) panels are the same as (a) and (b) except for the CH$_3$CN lines with two different upper energy levels.}
\label{fig:spec_mom2}
\end{figure*}

\begin{figure*}[hbt!]
\centering
\includegraphics[width=0.45\textwidth]{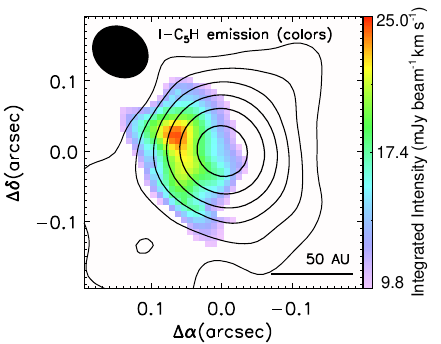}
\includegraphics[width=0.5\textwidth]{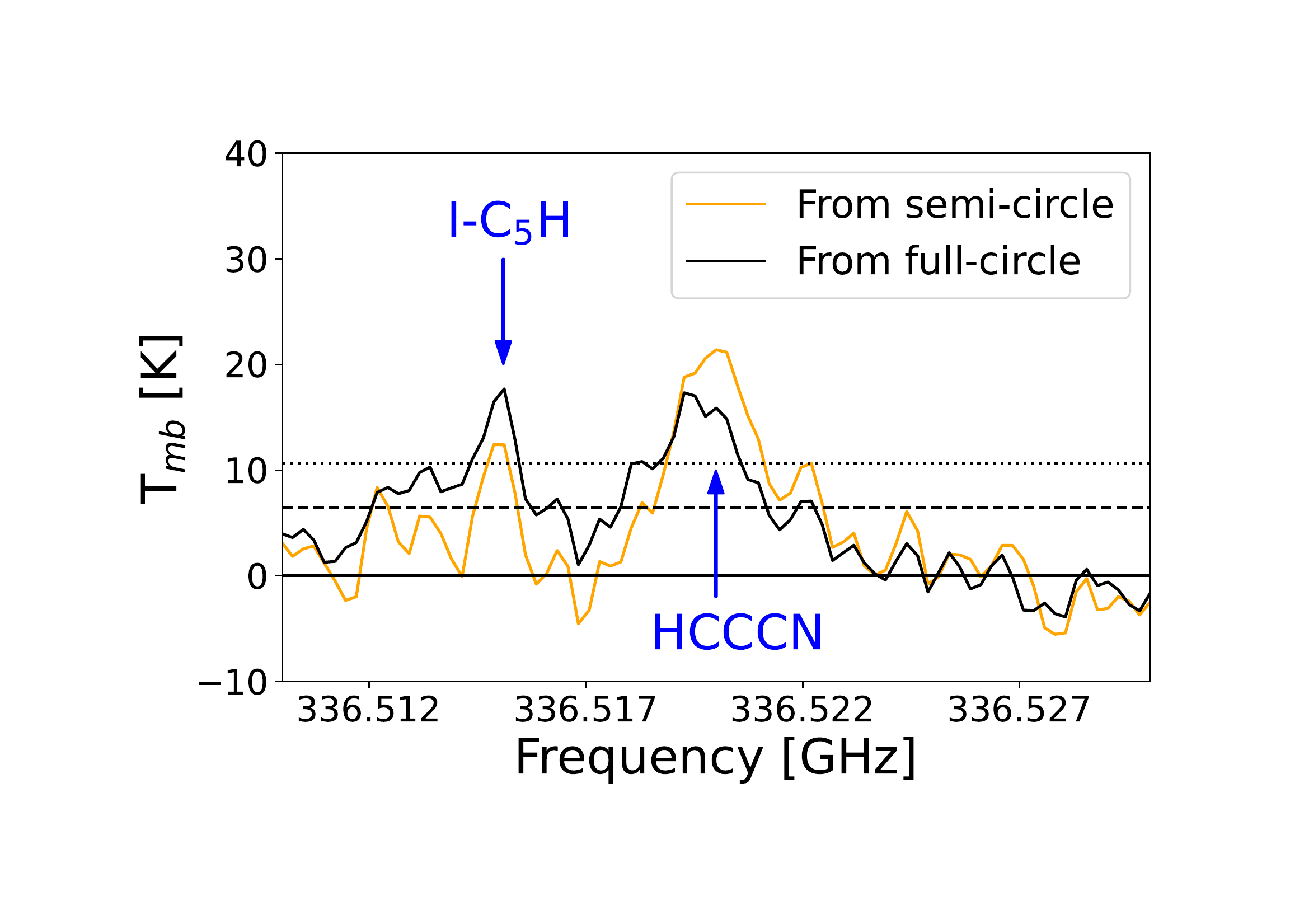}
\vspace{-1mm}
\caption{The integrated intensity map (color) of the C$_5$H line at 336.5148 GHz (left) and the spectra extracted from the full circle and the western semicircle as marked in black and orange colors, respectively, in Figure~\ref{fig:spec_mom2} (right). The emission distributes only in the eastern part of HOPS 373SW, where lines are affected by outflows/jets. Therefore, the C$_5$H line disappears in the spectrum extracted from the western semicircle (right). In the left panel, contours show the distribution of the dust continuum emission, and the levels are the same as those in Figure \ref{fig:cont_mol}. The integrated intensity over 5 $\sigma$ (1 $\sigma$ = 1.96 mJy beam$^{-1}$ km s$^{-1}$) is presented in the color image. The dashed and dotted lines in the right panel represent the 3 $\sigma_{T_{\rm mb}}$ and 5 $\sigma_{T_{\rm mb}}$ (1 $\sigma_{T_{\rm mb}}$ = 2.2 K) levels, respectively.}
\label{fig:c5h}
\end{figure*}

Even within 0.1\arcsec, the eastern hemisphere is largely affected by the outflow, particularly in the \chxoh\ lines. 
Figure \ref{fig:spec_mom2}(a) presents the line width distribution of the \chxoh\ and \xchxoh\ lines, and the bottom panel (b) compares the line profiles extracted from the regions with the same color boundaries. 
Unlike the \chxoh\ lines extracted from the sectors for the orange and red color boundaries tracing semicircles in the west, the \chxoh\ line extracted from the full annulus traced by the black color boundary shows a red-shifted high-velocity wing structure. 
This effect of outflow appears differently in different transitions of a given molecule.
A lower energy transition of CH$_3$CN (E$_u$=232 K) has a high-velocity dispersion in the eastern hemisphere as seen in the \chxoh\ emission, while its higher energy transition of E$_u$=424 K is not affected by the outflow (Panels (c) and (d) in Figure \ref{fig:spec_mom2}). 

Other complex molecular emission is also affected by the outflows.
The line at 336.5148 GHz that is clearly detected in the spectrum extracted over the full annulus disappears in the spectrum extracted from the western semicircle, as seen in the right panel of Figure \ref{fig:c5h}.
This line is possibly a C$_5$H line.
The integrated intensity map of the line is presented in the left panel of Figure \ref{fig:c5h}; the emission distributes only in the eastern part of HOPS 373SW, where the red-shifted outflow component contributes to the molecular excitation (Figure \ref{fig:spec_mom2}). 

\begin{figure*}[ht]
\centering
\includegraphics[width=\textwidth,trim=10mm 40mm 10mm 20mm]{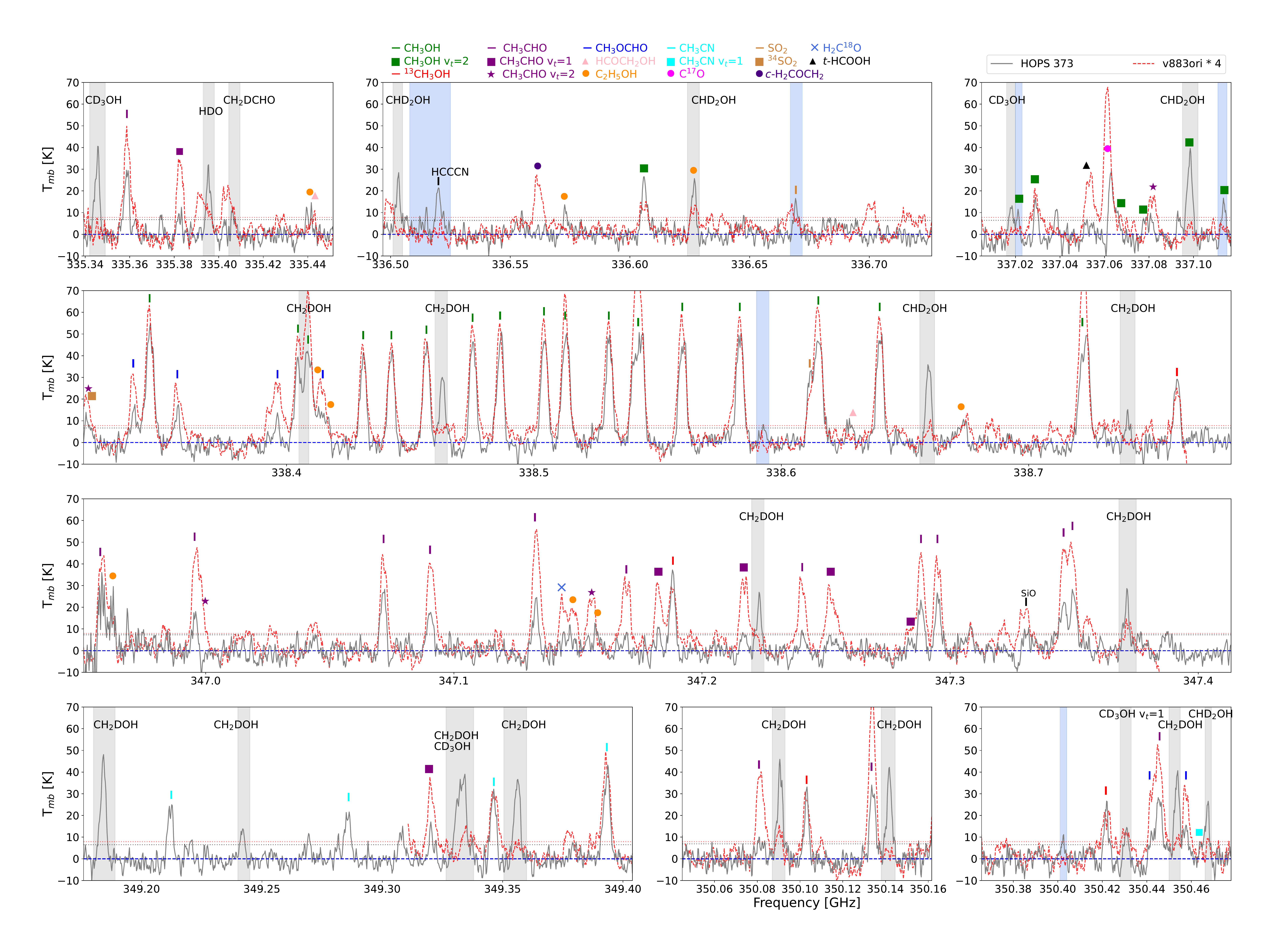}
\vspace{3mm}
\caption{The ALMA band 7 spectra of HOPS 373SW (gray solid line) compared with those of V883 Ori \citep[red dotted line, ][]{jelee19}. The spectral setup is almost the same in the two observations.
The spectra of V883 Ori are extracted over the full annulus between 0.1\arcsec to 0.2\arcsec\ in radius, while the spectra of HOPS 373SW are extracted over the western semicircle between 0.03\arcsec to 0.1\arcsec\ (see Figure~\ref{fig:spec_mom2}).
The V883 ori spectra have been scaled up by a factor of 4 to better compare with those of HOPS 373SW. The identified lines are marked with bars and symbols with different colors right above individual lines. The newly detected lines in HOPS 373SW and the lines much stronger than those of V883 Ori are shaded; the gray shades denote deuterated species, while the purple shades indicate unidentified or tentatively identified lines. The RMS noises ($\sigma_{T_{\rm mb}}$) are, before scaling up, 0.65 and 0.66 K for the spectral windows around 330 GHz and 340 GHz, respectively, in V883 Ori. In HOPS 373SW, the RMS noises ($\sigma_{T_{\rm mb}}$) are 2.2 and 2.3 K for the spectral windows around 330 GHz and 340 GHz, respectively. The red and gray dotted horizontal lines represent the 3 $\sigma_{T_{\rm mb}}$ level.}
\label{fig:spec_two}
\end{figure*}

The line profiles extracted from the sector with the red boundary, where all molecular emission is the strongest, are consistent with those extracted from the orange boundary sector, with similar line intensities. 
Therefore, we extract the spectra from the western semicircle inside the orange boundary to avoid the effect of the outflow and to maximize the S/N of each spectrum.
The extracted spectra from the western region outlined with the orange semicircle are used for 
the comparison with V883 Ori (Figure \ref{fig:spec_two}), the rotation diagram analysis (Figure \ref{fig:rot_dia}), and the detailed line fitting (Figure \ref{fig:spec_xclass}).



\section{Line Spectra}

\subsection{Comparison with V883 Ori}

The extracted spectra are presented in Figure \ref{fig:spec_two} in the gray line. 
These spectra were corrected for the velocity shifts caused by the rotational motion. We derived the rotation velocity structure using the \xchxoh\ lines, which is presented in the upper second panel of Figure \ref{fig:coms_m0m1} as the color image. The rotation motion in HOPS 373SW probably originates from the envelope rather than the disk (Section 5.1).
In fact, the infalling envelope is well traced by the inverse P-Cygni profile of C$^{17}$O J=2--1, which shows the red-shifted (i.e., at the lower frequency) absorption below the continuum (see the line profile marked with the magenta circle around 337.06 GHz in Figure~\ref{fig:spec_two}.)

The spectra of V883 Ori \citep{jelee19} are also plotted in Figure \ref{fig:spec_two} with the red dotted line for comparison. The spectra of V883 Ori were extracted over the annulus between 0.1\arcsec\ and 0.2\arcsec\ in radius and also corrected for the disk rotation \citep{jelee19}.
The \xchxoh\ line peak intensities of V883 Ori are about four times weaker than those of HOPS 373SW, so in Figure \ref{fig:spec_two}, the spectra of V883 Ori are scaled up by a factor of 4 for the comparison with the HOPS 373SW spectra. This lower intensity of V883 Ori compared to HOPS 373 is probably caused by a combination of higher dust continuum optical depth and lower gas temperature {\it at the emission peak beyond the emission hole} in the disk of V883 Ori ($\sim$0.15\arcsec) than in the envelope of HOPS 373 ($\sim$0.06\arcsec).

\begin{figure*}[t]
\centering
\includegraphics[width=0.3 \textwidth]{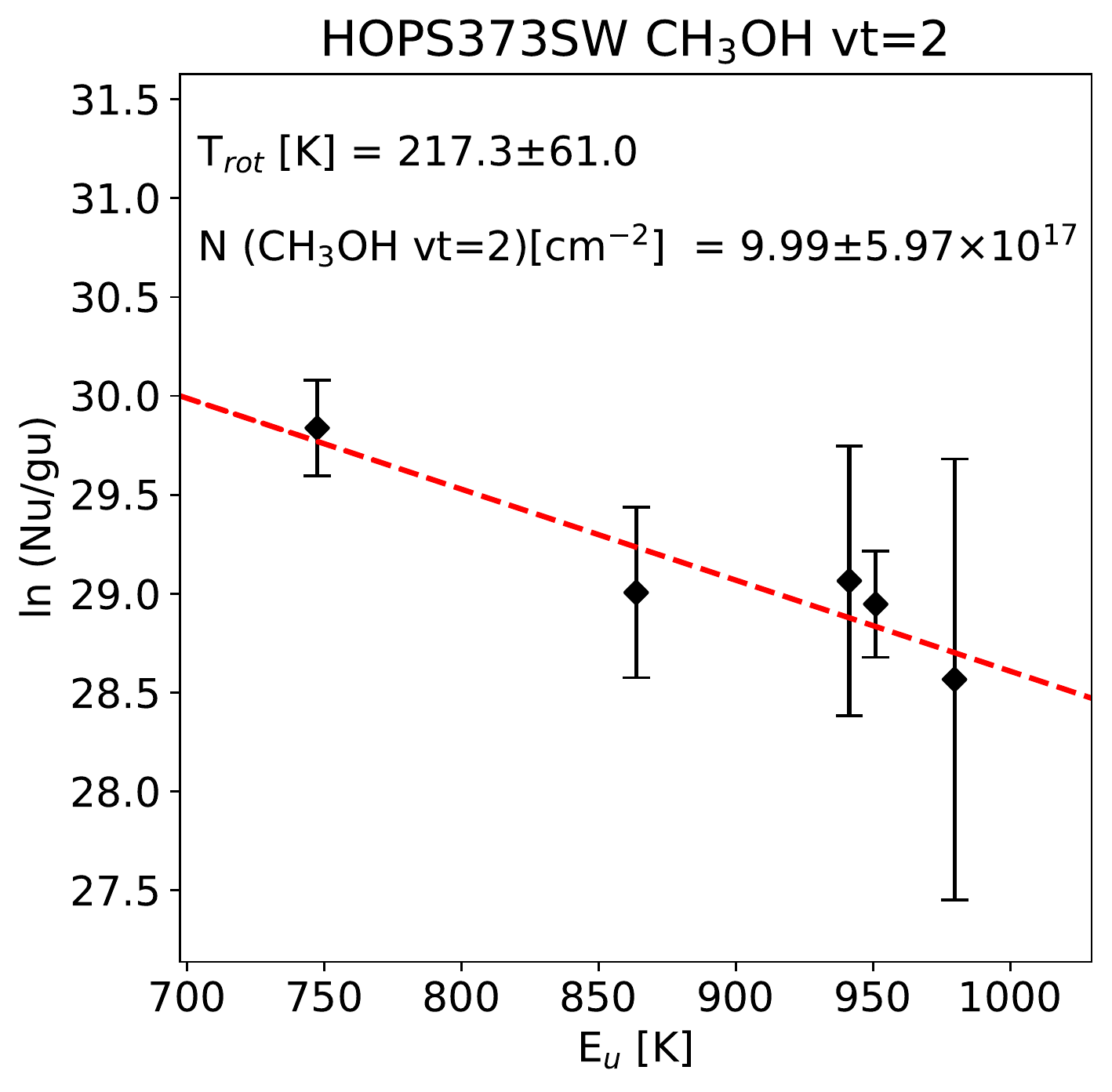}
\includegraphics[width=0.29 \textwidth]{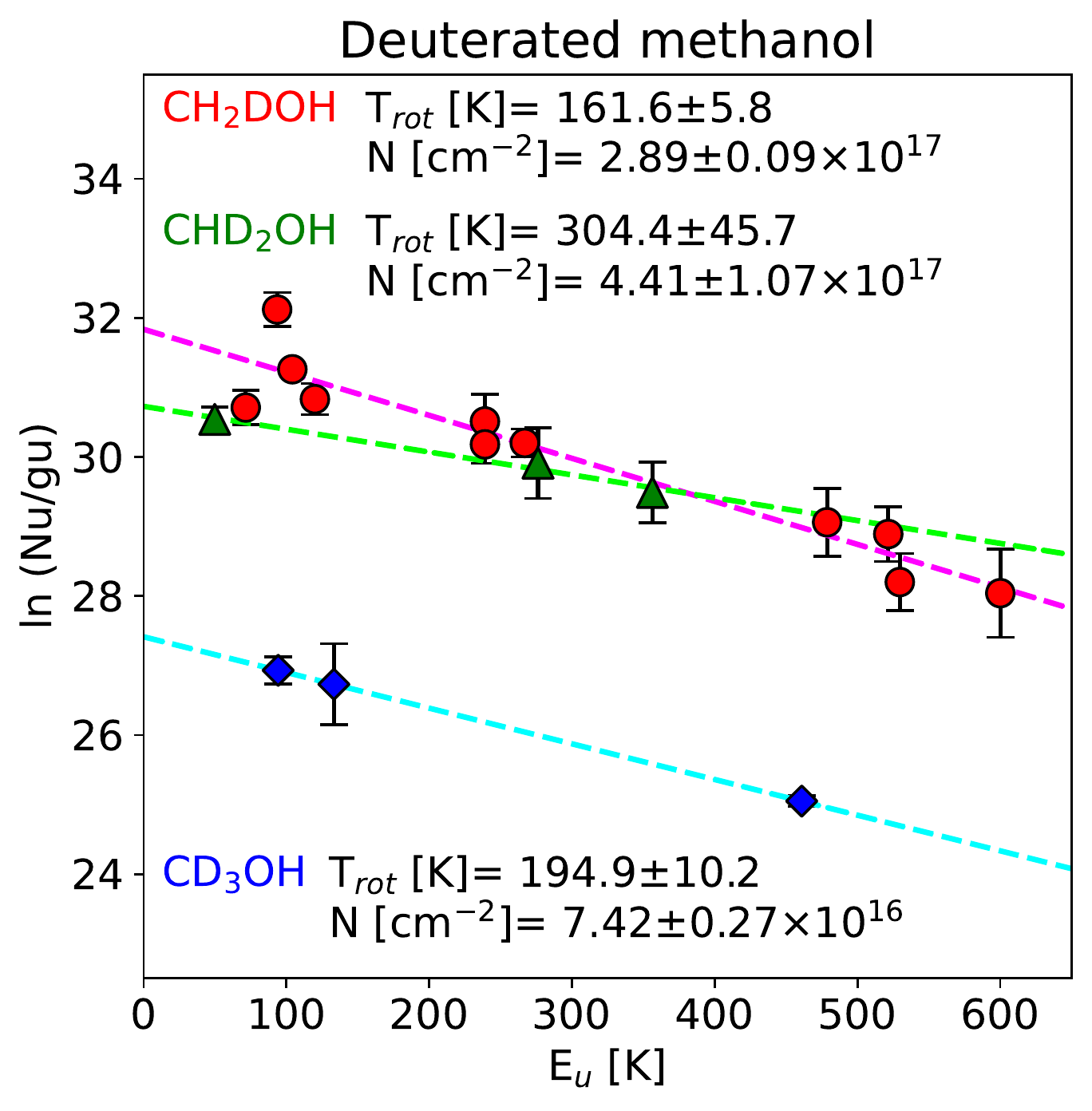}
\includegraphics[width=0.3 \textwidth]{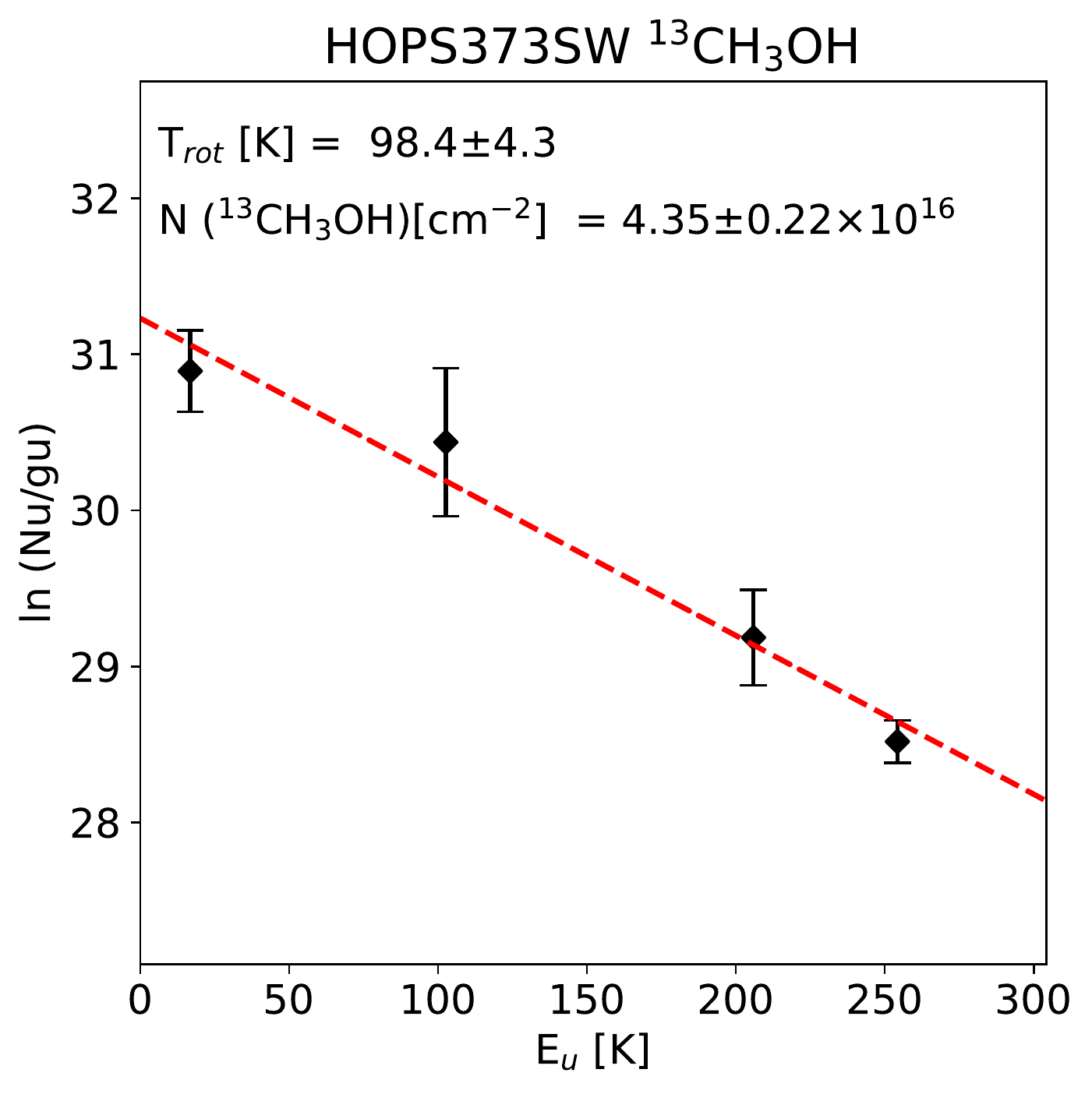}\\
\includegraphics[width=0.3 \textwidth]{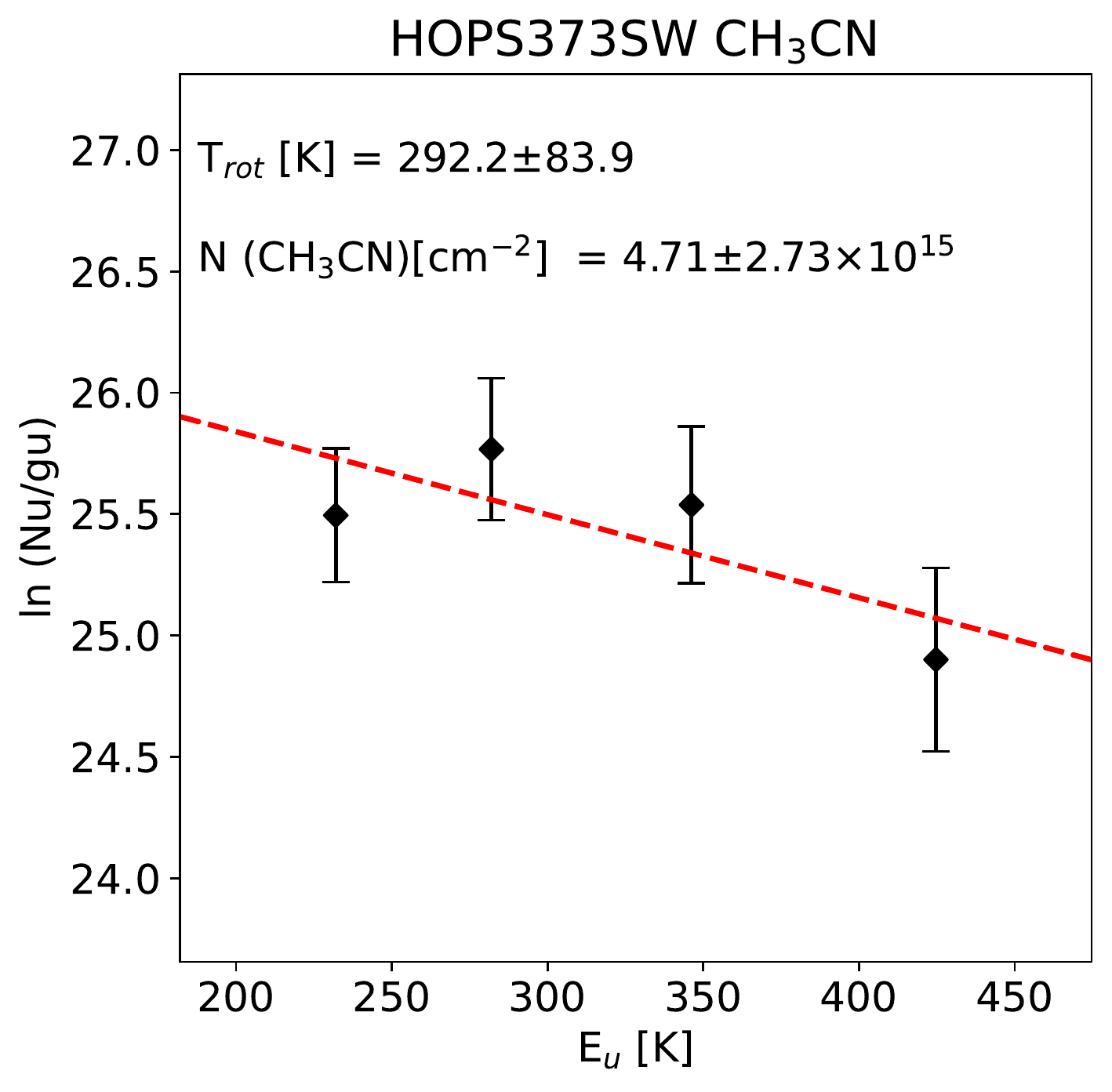}
\includegraphics[width=0.29 \textwidth]{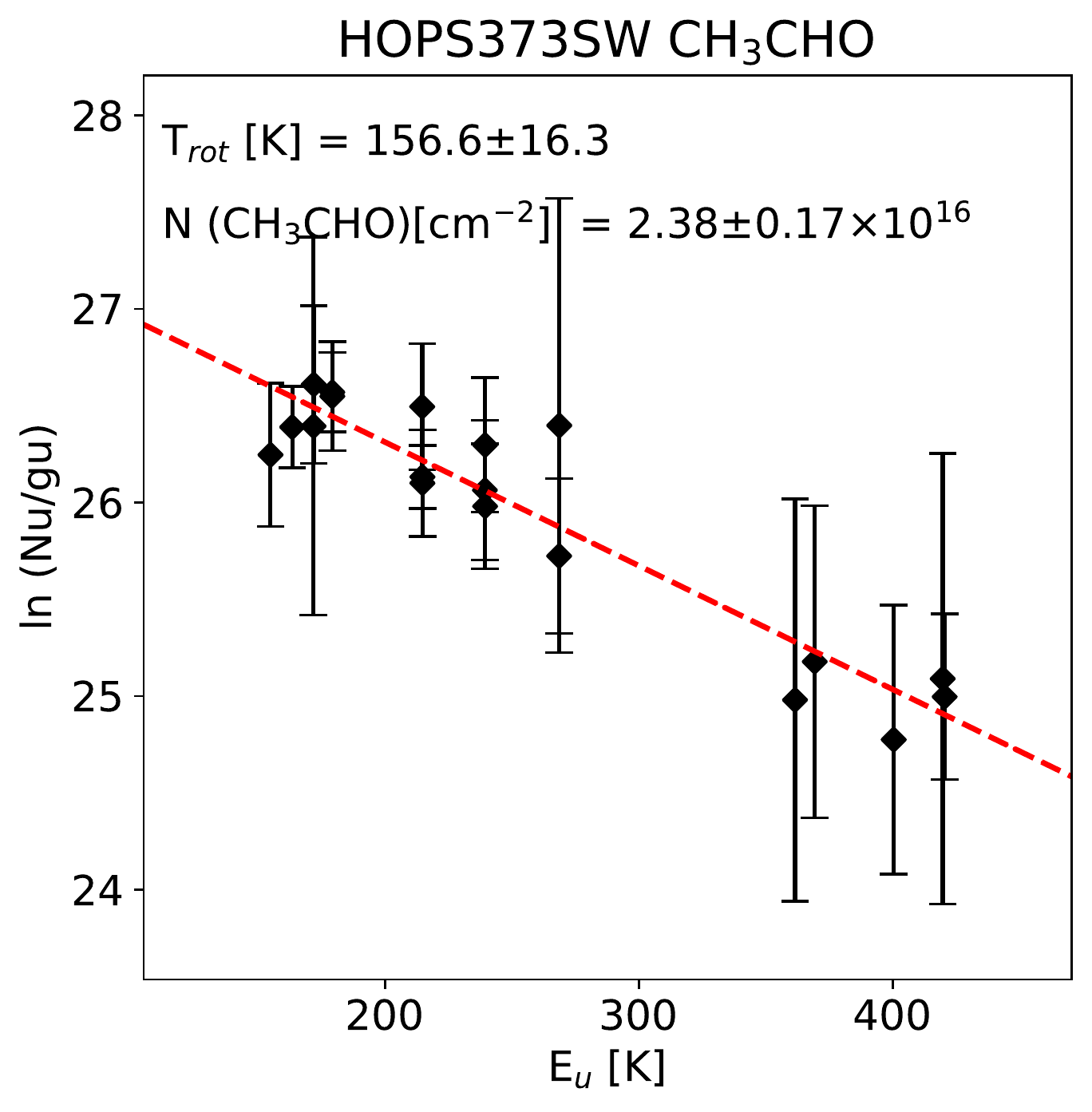}
\includegraphics[width=0.29 \textwidth]{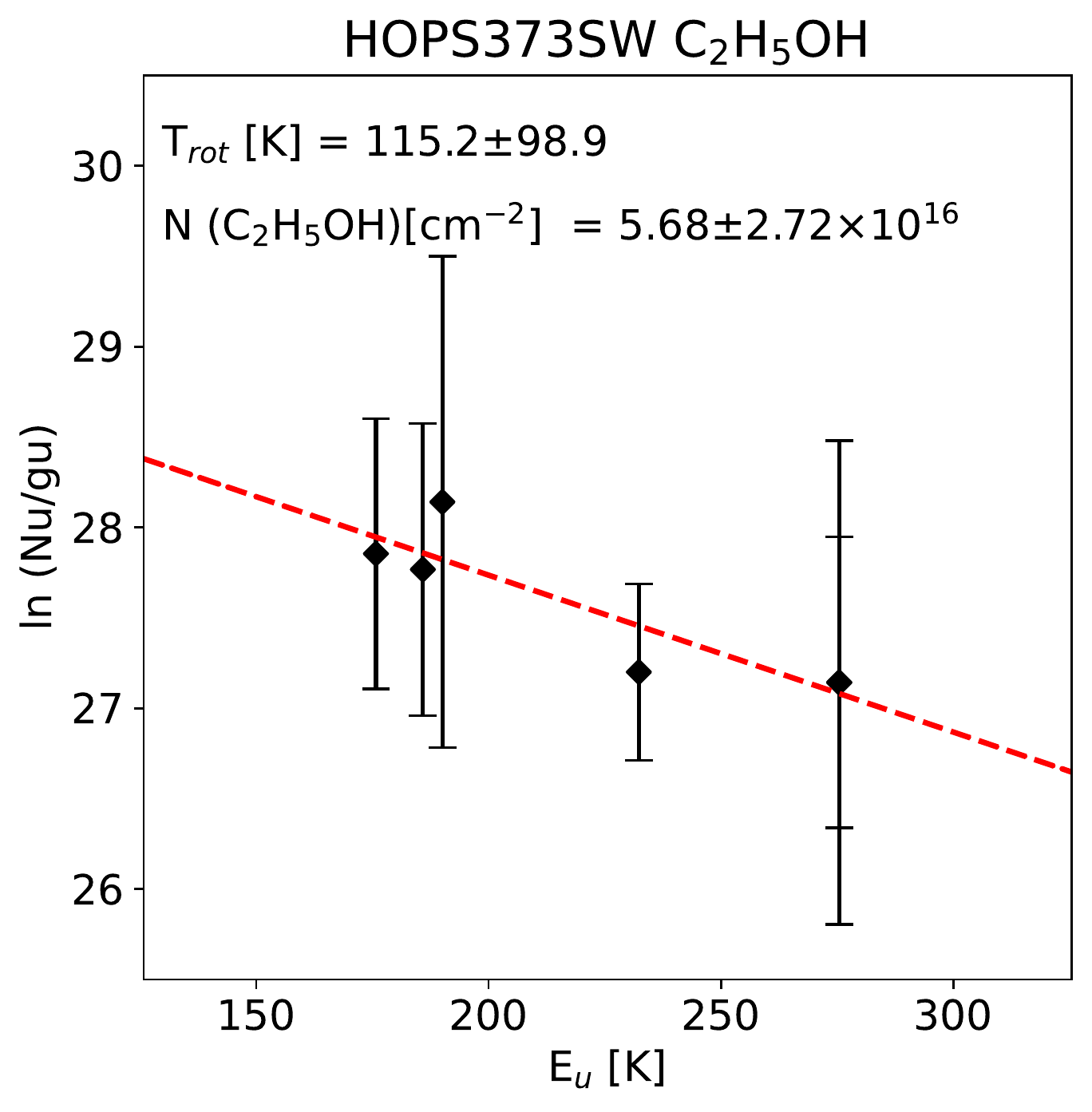}
\caption{The excitation diagrams of CH$_3$OH (v$_{t}$=2), CH$_{2}$DOH, CHD$_{2}$OH, CD$_{3}$OH, $^{13}$CH$_{3}$OH, CH$_3$CN, CH$_3$CHO, and C$_2$H$_5$OH. The isolated lines that can be fitted with a Gaussian profile are used for these excitation diagrams.}
\label{fig:rot_dia}
\end{figure*}

V883 Ori is an eruptive young star with a bolometric luminosity of $\sim$200 \lsun\ \citep{Furlan2016}, in transition from Class I to Class II, whereas HOPS 373SW is a very young Class 0 object with the bolometric luminosity of 5.3 \lsun. Therefore, one could expect that the chemical characteristics inherited from the natal molecular cloud have not been modified much in HOPS 373SW, while the chemistry in V883 Ori may have evolved significantly in the dense protoplanetary disk. 

From a simple comparison of the line intensities, appreciable differences between HOPS 373SW and V883 Ori are found, as follows: 
\begin{description}
\item[(1)] The HDO line is clearly detected in HOPS 373SW without any confusion by neighboring lines, while it is marginally detected in V883 Ori.
\item[(2)] The C$^{17}$O line shows red-shifted absorption against the continuum in HOPS 373, indicative of infall in the inner envelope.
\item[(3)] The \chxoh\ v$_{\rm t}$=2 lines have {\bf only} been detected in HOPS 373, suggesting a higher gas temperature than in V883 Ori.
\item[(4)] Many strong CH$_2$DOH, CHD$_2$OH, and CD$_3$OH lines are detected in HOPS 373SW. The lines are as strong as the main isotopic methanol lines, indicative of a very high D/H ratio in methanol even in consideration of the optically thick CH$_3$OH lines.
\item[(5)] The CH$_3$CHO line intensity relative to that of \chxoh\ is weaker in HOPS 373SW than in V883 Ori, whereas one line of CH$_2$DCHO is detected at 335.406 GHz with a similar strength to the CH$_3$CHO v$_{\rm t}$=1 line at 335.382 GHz in HOPS 373SW.
\item[(6)] The CH$_3$OCHO line intensity relative to that of \chxoh\ is mostly weaker in HOPS 373SW than V883 Ori.
\item[(7)] The C$_2$H$_5$OH lines are detected clearly in HOPS 373SW, unlike in V883 Ori.
\item[(8)] The c-H$_2$COCH$_2$ and t-HCOOH lines, which are detected in V883 Ori, are not detected in HOPS 373SW.
\end{description}

\begin{figure*}[htb]
\centering
\includegraphics[width=1\textwidth]{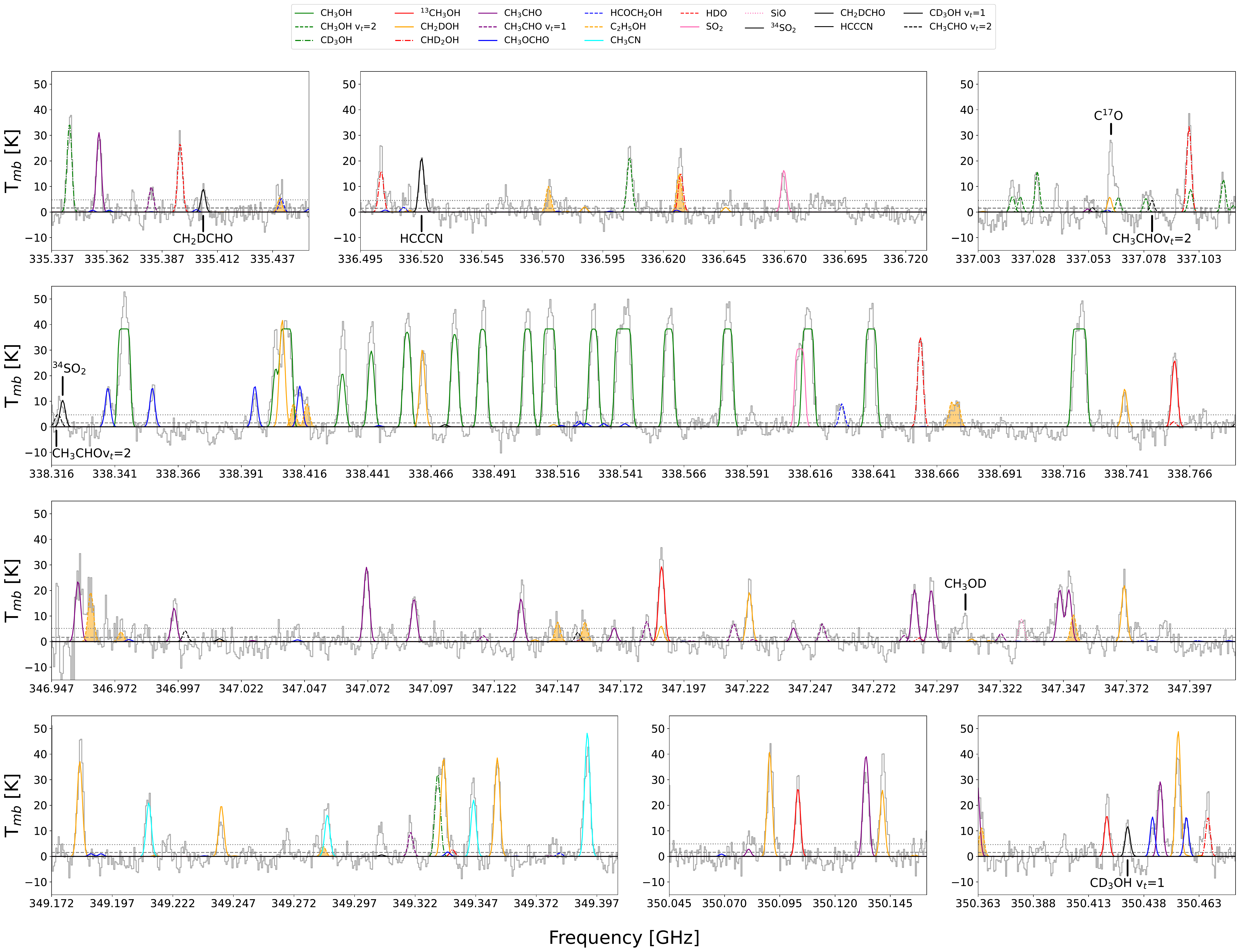}
\caption{The spectra of HOPS 373SW fitted with XCLASS. Different colors represent the line transitions of different species. Black solid/dashed lines represent the tentatively identified lines. C$_{2}$H$_{5}$OH lines are largely blended with other species. Thus, to clearly distinguish, we shaded C$_{2}$H$_{5}$OH lines with an orange color. The identified lines are listed in Table \ref{tb:identified_lines} and Table \ref{tb:tentative_detection}, and the column density and excitation temperature best fitted for each species are summarized in Table \ref{tb:xclass_results} along with its abundance relative to CH$_3$OH. The dashed and dotted horizontal lines represent the 1 $\sigma_{T_{\rm mb}}$ and 3 $\sigma_{T_{\rm mb}}$ levels, respectively.}
\label{fig:spec_xclass}
\end{figure*}

\subsection{The LTE analysis of COMs lines}

The excitation diagrams of eight species are presented in Figure \ref{fig:rot_dia}.
Compared to those of V883 Ori \citep[$\sim$100 K]{jelee19}, the excitation temperatures are very high; the excitation temperatures of CH$_2$DOH and CH$_3$CHO are $\sim$160 K, and those of CHD$_2$OH and CH$_3$CN  
are even higher $\sim$300 K. HH 212 also shows high excitation temperatures of 150$\sim$160 K \citep{cflee19}. On the other hand, in most of ALMASOP hot corinos, the excitation temperatures of CH$_2$DOH are lower than 100 K \citep{Hsu22}. 

For more detailed analyses, we adopt XCLASS, which can identify the detected lines and derive the column densities and the excitation temperatures of species by inspecting all transitions of a species within the observed frequency range. 
We first fitted the line transitions of individual species independently to identify detected transitions and derive rough estimations of fitting parameters such as column density, excitation temperature, and line width for a specific molecule. 
Then, we fitted simultaneously ten molecules (CH$_{3}$OH v$_{t}$=2, CD$_{3}$OH, $^{13}$CH$_{3}$OH, CH$_{2}$DOH, CHD$_{2}$OH, CH$_3$CHO, CH$_3$OCHO, C$_2$H$_5$OH, CH$_3$CN, HCOCH$_{2}$OH), whose lines are heavily blended, especially in the 4th and 8th spectral windows, to provide the final column densities and excitation temperatures listed in Table \ref{tb:xclass_results}. The other molecules are fitted individually. 
Figure \ref{fig:spec_xclass} shows the spectra fitted by XCLASS on top of the observed spectra.
Before the XCLASS fittings, lines are corrected for the velocity shifts caused by rotation, and the central velocity is shifted to zero. 

XCLASS considers the dust continuum optical depth in the line fitting process using a given value of the molecular hydrogen column density. An optical depth ($\tau_c$) for the dust continuum emission reduces the molecular line intensity by $\exp(-\tau_c)$ \citep{Yen2016,jelee19,shlee20}.
We derive the average  $\tau_c$ over the western semicircle marked by the orange color boundary in Figure \ref{fig:spec_mom2}, where the line spectra were extracted, using a simple model. 
We assumed that the dust temperature distribution follows T$_{0.03\arcsec} \times \sqrt{0.03\arcsec/r}$, where T$_{0.03\arcsec}$ is the dust temperature at r$=$0.03\arcsec, and dust emission is very optically thick near the continuum peak; then, the observed continuum peak could be reproduced with T$_{0.03\arcsec} \sim$105\ K. Median values of the dust continuum intensity (J(T)) and the model dust temperature  over 0.03\arcsec\ to 0.1\arcsec\ in the western semicircle (outlined with the orange boundary) are 34~K and $\sim$70 K, respectively, which result in $\tau_c$ of 0.66. Finally, a molecular hydrogen column density of $\sim$8.9 $\times 10^{24}$ cm$^{-2}$ is derived from $\tau_c$ by assuming the OH5 dust opacity $\kappa_\nu =$ 1.84 cm$^{2}$~g$^{-1}$ \citep{Ossenkopf94}, the gas-to-dust ratio of 100, and the mean molecular weight of 2.4.

The column densities derived from the XCLASS fitting are generally higher than the values derived from the rotation diagrams, except for CH$_3$CN. However, considering the uncertainties, they are consistent within a factor of 2, at maximum. Typically, the derived temperatures are also consistent at the level of a few 10s of degrees although those of CHD$_2$OH, CD$_3$OH, and CH$_3$CN show rather large differences of 70 to 90 K. We believe that the values from the XCLASS fitting are more reliable because the dust opacity, as well as self-absorption, has been taken into account by XCLASS.



\section{Discussion}

\subsection{The origin of COMs emission: hot corino vs. disk}

\begin{figure}[t]
\centering
\includegraphics[trim=0cm 0.3cm 0cm 0cm, clip=true,width=0.47\textwidth]{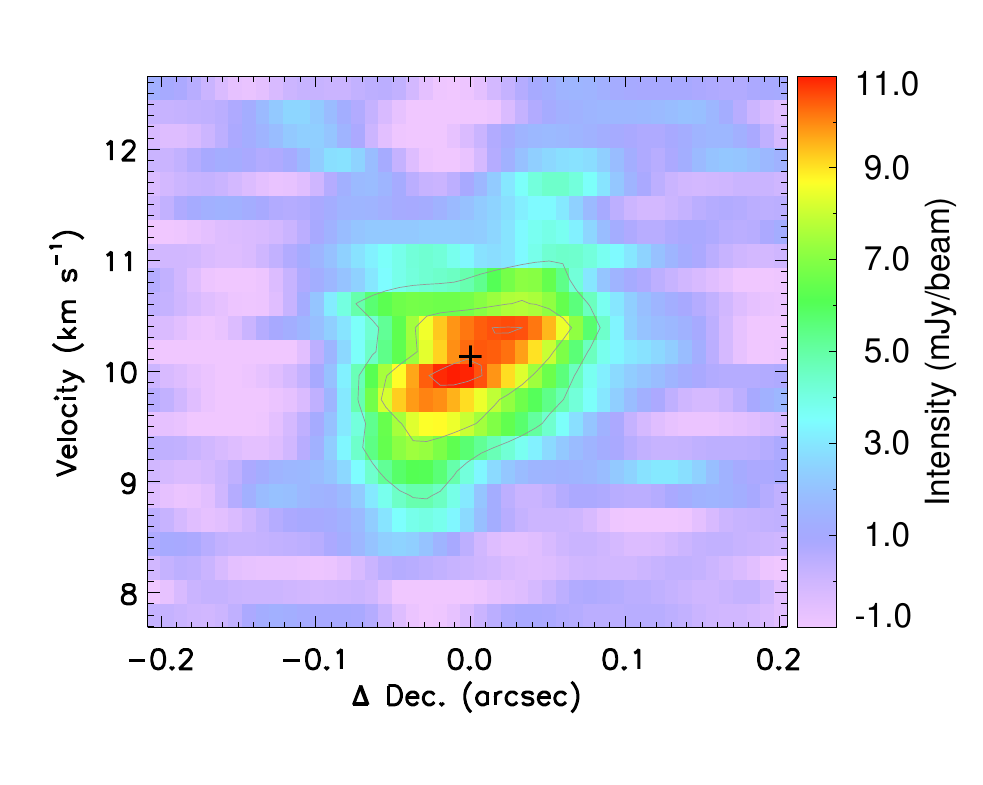}
\vspace{-0.8cm}
\caption{Position velocity diagram of \xchxoh\, along the Declination. The cross indicates the source velocity (10.13 \kms) and the continuum peak position. The contours are 5, 8, 11 $\sigma$  (1 $\sigma$= 1.8 \funit). }
\label{fig:pv}
\end{figure}

The clear velocity structure of rotation is shown in the intensity weighted velocity maps of most detected simple or complex molecules (see examples in Figure \ref{fig:coms_m0m1}). In the early evolutionary stage, particularly in the extremely young Class 0 object, the kinematics may be dominated by the envelope rather than the disk even at this sub-arcsecond resolution since the disk may not be well developed yet. The position-velocity diagram of a $^{13}$CH$_3$OH line (Figure \ref{fig:pv}) does not show a Keplerian disk rotation feature, although a detailed examination of the kinematics, beyond the scope of this paper, is required to confirm the real origin of COMs emission. Nevertheless, as we will see from the emission radius, we speculate that the detected rotational motion in HOPS 373SW is likely of the dense inner envelope, and the disk radius could be much smaller than the radius of continuum emission ($\sim$65 au). 

In general, the hot corino sources with rich COMs emission, such as IRAS 16293B \citep{Jorgensen2016}, B335 \citep{Imai16}, and L483 \citep{Jacobsen19}, do not have well-developed large disks. In contrast, protostars with well-developed massive disks with sizes $\sim$100 au \citep{Sakai14, villarmois19, shlee20, Hoff20} lack rich COMs emission, suggesting that COMs are frozen on grain surfaces in the dense disk midplane since the current luminosity of those protostars is not high enough to extend the water snow line to a large radius in the disk midplane. 
In this sense, the rich COMs emission detected in HOPS 373SW very likely originates from the hot corino, the hot inner envelope. 

The water sublimation radius can be another discriminator for the origin of COMs emission.
The water sublimation radius in a hot corino can be estimated from the source bolometric luminosity \citep{Bisschop07, van'tHoff22}. According to the 1-D radiative transfer models of massive protostars by \citet{Bisschop07}, the water sublimation radius follows the equation, 
$R_{\rm snow} \sim R(100~{\rm K}) = 15.4 \sqrt{L/\lsun}$ au. 
This relation can also be applied to low-mass protostars \citep{van'tHoff22}.
With the luminosity of 5.3 \lsun, from the above equation, the water sublimation radius of HOPS 373SW is anticipated as $\sim$35 au, which is consistent with the boundary ($\sim$0.1\arcsec) of COMs emission. This consistency also supports that the COMs emission originates from the hot corino rather than a disk in HOPS 373SW. 

The eruptive disk source, V883 Ori, has a bolometric luminosity of $\sim$200 \lsun, which should develop the water sublimation radius around 220 au in a hot corino. However, the estimated snow line in V883 Ori is located between 40 au and 100 au \citep{jelee19, leemker21, tobin23}. This indicates that the relation between water sublimation radius and luminosity is different in hot corinos and disks, probably due to their different densities and dust properties \citep{Chang97}.



\subsection{Deuteration of \chxoh\ and CH$_3$CHO}
The most remarkable chemical characteristics in HOPS 373SW are very high D/H ratios of methanol and the strong HDO and CH$_2$DCHO lines. 
The detection of even doubly and triply deuterated methanols has been confirmed in HOPS 373SW, thanks to the newly calculated spectroscopic data \citep{Coudert21, Ilyushin22}.
Table~\ref{table:deut} lists the column density ratios of deuterated species as well as the D/H ratios corrected for the statistical effect \citep{Parise04}. 

In Table~\ref{table:deut}, the column density ratio between CH$_2$DOH and CH$_3$OH is 0.36. The ratio can be 0.13 
if we use the \xchxoh\ column density and the carbon isotopic ratio ($^{12}$C/$^{13}$C) of 60 \citep{jelee19}.
However, many \chxoh\ v$_{\rm t}$=2 lines, which are optically thin, have been detected in HOPS 373SW. Therefore, for the calculation of the D/H ratio, we adopt the column density derived from the v$_{\rm t}$=2 transitions, which is, in fact, not very different from what has been derived from the v$_{\rm t}$=0 lines with XCLASS. 
In addition, the excitation temperature derived from the v$_{\rm t}$=2 lines is similar to those of deuterated methanol.
The $^{12}$C/$^{13}$C ratio is about 20 if we use the \chxoh\ column density derived from the \chxoh\ v$_{\rm t}$=2 lines and the \xchxoh\ column density. This low value is consistent with the value derived in IRAS 16293B \citep{Jorgensen18}.

\begin{figure*}[htp]
\centering
\includegraphics[trim=0cm 0.5cm 0cm 0cm, clip=true,width=0.7\textwidth]{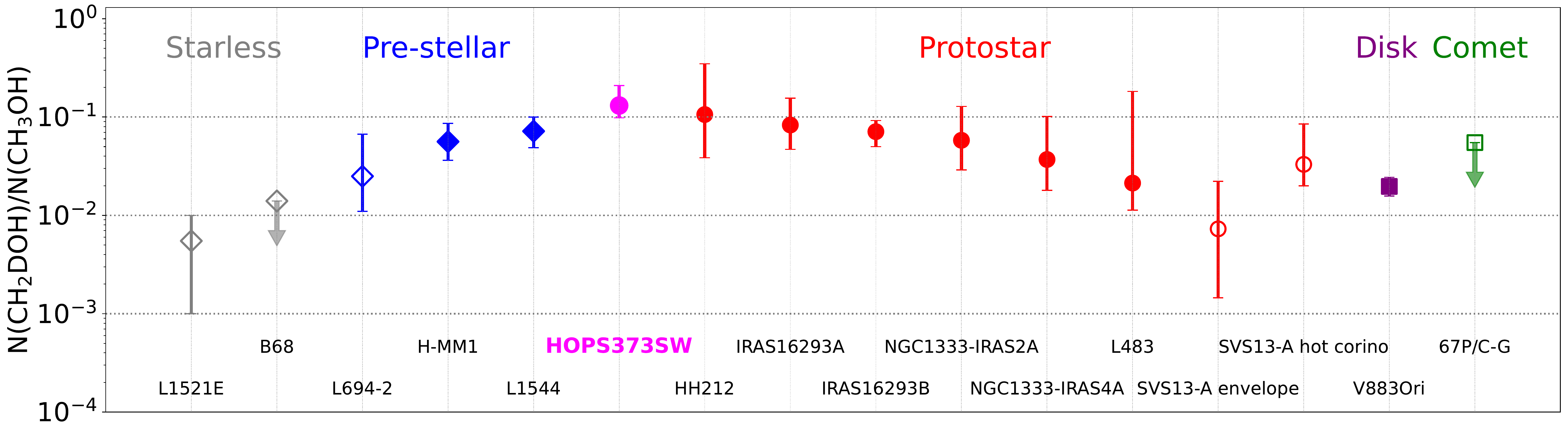}
\includegraphics[trim=0cm 0.5cm 0cm 0cm, clip=true,width=0.7\textwidth]{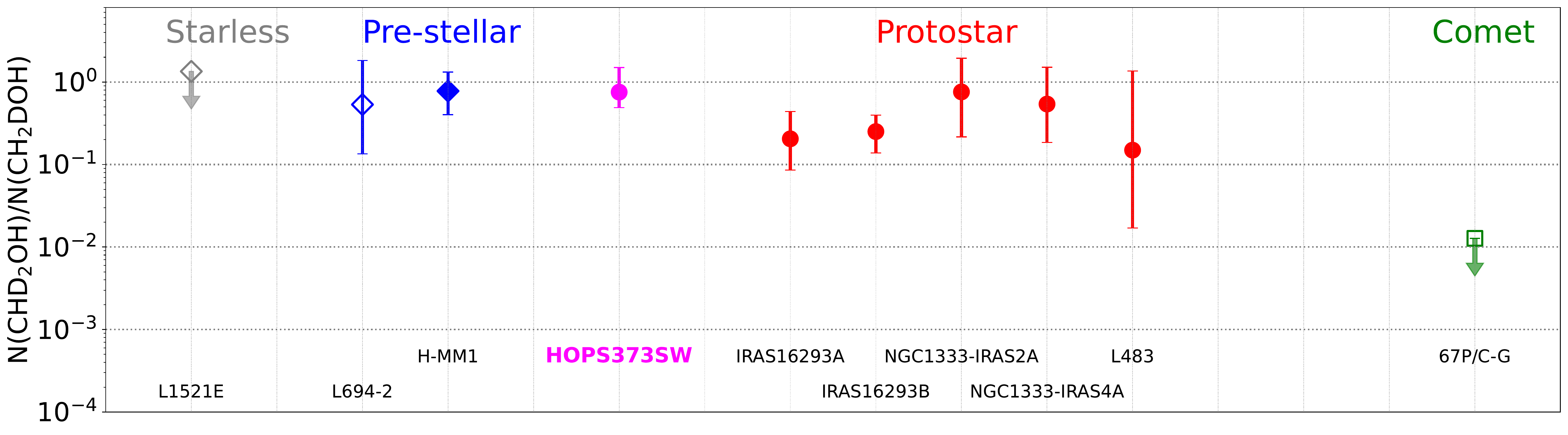}
\caption{Column density ratios of CH$_{2}$DOH/\chxoh~(upper panel) and CHD$_{2}$OH/CH$_{2}$DOH (lower panel). Filled and open symbols indicate single-dish and interferometric observations, respectively. Except for HOPS 373SW, data have been obtained from the literature; L1521E, B68, L694-2, and H-MM1 from \citet{Lin23}, L1544 from \citet{Chacon-Tanarro2019A&A...622A.141C}, HH212 from \citet{cflee19}, IRAS16293A and IRAS16293B from \citet{Jorgensen18}, \citet{Manigand2019} , and \citet{Drozdovskaya2022}, NGC1333-IRAS2A and NGC1333-IRAS4A from \citet{Taquet2019A&A...632A..19T}, L483 from \citet{Agundez2019A&A...625A.147A}, SVS13 from \citet{Bianchi2017MNRAS.467.3011B}, V883Ori from \citet{jelee19}, and 67P/C-G from \citet{Drozdovskaya2021MNRAS.500.4901D} }
\label{fig:d_to_h_ratio}
\end{figure*}

HOPS 373SW is the second hot corino where the detection of CD$_3$OH has been confirmed, following IRAS 16293B \citep{Ilyushin22}. For HOPS 373SW, the column density ratio between CHD$_2$OH and CH$_2$DOH is 0.75, which is much higher than the ratio between CH$_2$DOH and CH$_3$OH. This trend is consistent with other hot corinos \citep{Drozdovskaya2022} and prestellar cores \citep{Lin23}. Figure~\ref{fig:d_to_h_ratio} shows that the column density ratios of singly and doubly deuterated methanols in HOPS 373SW are higher than other hot corinos and similar to the highest values seen in prestellar cores.

A transition of CH$_{3}$OD, another singly deuterated methanol, is covered in our spectral setup. An emission line is clearly detected with $>$ 3$\sigma$ (T$_{\rm mb}$ of $\sim$10 K) at 347.308432 GHz \citep{Anderson1988ApJS...67..135A,Duan2003JMoSp.218...95D}. This tentative detection of CH$_{3}$OD also supports the high deuteration of HOPS373 SW since the deuteration in the OH functional group is less efficient compared with the CH$_{3}$ functional group \citep{Nomura2022arXiv220310863N}.

As seen in Table~\ref{table:deut}, the D/H ratios with the statistical corrections are higher for multi-deuterated methanols than the mono-deuterated methanol by a factor of $\sim$3, suggesting that subsequent hydrogen abstractions may be more probable after the first deuteration of methanol \citep{Drozdovskaya2022}.
According to the chemical models of deuterated ices by \citet{Taquet14}, the D/H ratio of deuterated methanol can be enhanced if a hot corino had a higher density in the prestellar stage.
In addition, the D/H ratios of singly deuterated methanol and singly deuterated acetaldehyde are similar to $\sim$10 \%, which is the highest level for hot corinos \citep{Drozdovskaya2022}.
This enhanced deuteration, which is believed to occur in the cold prestellar phase \citep{jelee15}, strongly suggests that HOPS 373SW is a very young hot corino that retains the memory of the cold phase since the D/H ratio decreases with protostellar evolution \citep{aikawa12}. 
In addition, the similar D/H ratio of methanol and acetaldehyde indicate that the two molecules formed over a similar timescale and/or within a similar layer of the ices \citep{Taquet14}.

\begin{figure}[ht!]
\centering
\includegraphics[width=0.47\textwidth]{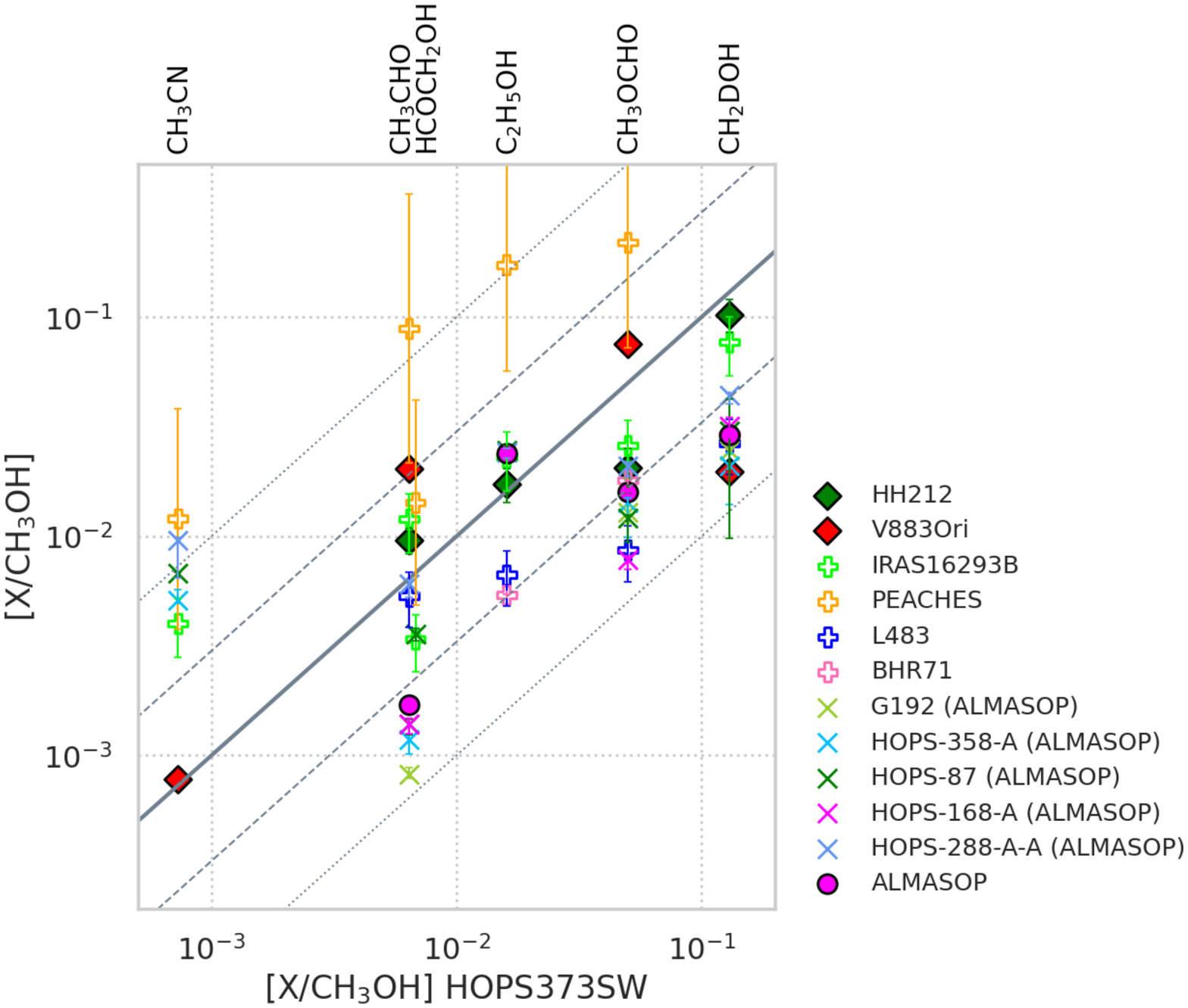}
\caption{Comparison of COMs abundances with other hot corinos and disks. The abundances toward HOPS 373SW and other sources (Table \ref{tb:comparison}) are on the x- and y-axis, respectively. Some abundances from the literature are rescaled to be consistent with our assumption of $^{12}$C/$^{13}$C=60. The diagonal lines denote equal abundances (solid) and abundances factors of 3 (dashed) and 10 (dotted) above or below those in HOPS 373SW. }
\label{fig:abun_comp}
\end{figure}

\subsection{COMs abundances vs. evolution}

The region of COMs emission in HOPS 373SW is about 40 au, which is consistent with the water sublimation radius developed in the envelope by its luminosity \citep{Bisschop07, van'tHoff22} as discussed in Section 5.1.
Table \ref{tb:comparison} and Figure \ref{fig:abun_comp} compare the abundances of COMs detected in HOPS 373SW with those of other hot corinos in Orion \citep[ALMASOP; ][]{Hsu22}, HH 212 \citep{cflee19}, and V883 Ori \citep{jelee19} as well as other well studied hot corinos, including IRAS 16293B \citep{Jorgensen2016} and the Perseus ALMA Chemistry Survey (PEACHES) samples \citep{Yang21}. We use the $^{13}$CH$_3$OH lines to derive the COMs abundances respective to CH$_3$OH for consistency with other studies.
In general, $^{13}$CH$_3$OH or CH$_3$$^{18}$OH has been adopted to calculate the column density of  CH$_3$OH because the CH$_3$OH lines are optically thick. According to our XCLASS fitting results, the optical depths are as high as $\sim$40 for some transitions of \chxoh, while the optical depths of all \xchxoh\ transitions are below 1.

The abundances of the ALMASOP hot corinos and HH 212, which were obtained from the literature, were scaled down by a factor of 1.2 in Table \ref{tb:comparison} because they were originally calculated by adopting the $^{12}$C/$^{13}$C ratio of 50 \citep{Kahane2018}, while we use the ratio of 60 \citep{Langer1993} for HOPS 373SW and V883 Ori. 
Note, however, that the true $^{12}$C/$^{13}$C may be about 20 for HOPS 373SW (see Section 5.1) and potentially V883 Ori (Jeong et al. in prep.), much lower than 60.
In addition to the mean abundances taken over 11 ALMASOP sources, we also list the COMs abundances of 5 individual hot corinos, which have the richest COMs spectra among the ALMASOP sample, separately at the bottom of Table \ref{tb:comparison}.

In general, the PEACHES sample (orange cross in Figure \ref{fig:abun_comp}) has significantly higher COMs abundances than HOPS 373SW. 
The most diverse abundance distribution is found for CH$_3$CN and CH$_3$CHO. 
In all other hot corinos, 
compared to HOPS 373SW, CH$_3$CN is much more abundant, while other COMs such as HCOCH$_2$OH and CH$_3$OCHO are less abundant. The ratio between CH$_3$CN and \chxoh\ has been suggested as an evolutionary indicator \citep{Oberg14}, since the formation of CH$_3$CN on grain surfaces is sensitive to the physical conditions and evolutionary timescales in individual sources \citep{Bonfand19}. Therefore, the low ratio of CH$_3$CN relative to \chxoh\ suggests that HOPS 373SW is younger than other hot corinos.

HOPS 373SW, HH 212, and V883 Ori as well as ALMASOP hot corinos are located in the Orion Molecular Cloud Complex, so their initial chemical compositions may be similar. Therefore, the chemical differences between these sources could be considered as the consequence of evolution from envelope to disk.

Compared with the ALMASOP hot corinos, HOPS 373SW has a very high D/H ratio ($> 0.1$) of \chxoh, a lower abundance of CH$_3$CN, and a higher abundance of HCOCH$_2$OH.
On the other hand, HOPS 373SW shows very similar abundances to HH 212, except for CH$_3$OCHO, whose abundance is higher only by a factor of $\sim$2 in HOPS 373SW than HH 212.
This may suggest that COMs did not have enough time to be altered in HH 212, and the physical conditions in the disk surface of a Class 0 protostar are not very different from the inner envelope exposed to the radiation from the central Class 0 protostar. 
However, in V883 Ori, the D/H ratio of \chxoh\ is reduced by a factor of 6.5, while the CH$_3$CHO abundance is enhanced by a factor of 3, compared to HOPS 373SW, although CH$_3$CN 
has similar abundance in both sources. 
The D/H ratios of water and COMs delivered to the protoplanetary disk could be reduced by turbulent mixing as suggested by \citet{furuya13}.

\section{Conclusion}

We carried out an ALMA observation of HOPS 373SW, which is a brightened and very embedded Class 0 protostar, in band 7, with the same spectral setup used for V883 Ori, which is an eruptive Class I/II disk source. This brightening event provides a unique opportunity to detect the COMs emission lines by expanding the sublimation radius due to the increased luminosity. In addition, comparing COMs emission between two distinct evolutionary stages in the same formation environment is critical to understanding chemical evolution in the grain surfaces.  In HOPS 373SW, eleven COMs as well as HDO and HCCCN have been detected. A rotation velocity structure is well resolved in most of the detected lines although a detailed examination of the kinematics needs to confirm the origin of the rotation. Nevertheless, the size of COMs emission is consistent with  the {\it hot corino} snow line location expected from its luminosity. 

According to the comparisons with V883 Ori and other hot corinos, HOPS 373SW shows a highly enhanced D/H ratio of methanol, indicative of a very young hot corino stage. Even doubly and triply deuterated methanols are detected. The ratio between CH$_3$CN and \chxoh, which is an evolutionary indicator, is also low in HOPS 373SW, compared to other hot corinos. 
These chemical characteristics are well consistent with the nature of HOPS 373SW classified as a PACS Bright Red Sources (PBRS) \citep{Stutz2013}, which is considered extremely young. Therefore, HOPS 373SW could be a good place to study the chemistry of COMs in the cold prestellar phase, using the freshly sublimated COMs from grain surfaces.

\section{Acknowledgements}

The authors thank the referee for a helpful and careful report. This work was supported by the National Research Foundation of Korea (NRF) grant funded by the Korea government (MSIT) (grant number 2021R1A2C1011718).  G.J.H.\ is supported by general grants 12173003 and 11773002 awarded by the National Science Foundation of China.
D.J.\ is supported by NRC Canada and by an NSERC Discovery Grant. J.J.T. acknowledges support from  NSF AST-1814762. G.B. is supported by SNU Junior Research Fellowship through Seoul National University.
The National Radio Astronomy Observatory is a facility of the National Science Foundation operated under cooperative agreement by Associated Universities, Inc.

This paper makes use of the following ALMA data: ADS/JAO.ALMA\#2019.1.00386.T.
ALMA is a partnership of ESO (representing its member states), NSF (USA) and NINS (Japan), together with NRC (Canada), NSC and ASIAA (Taiwan), and KASI (Republic of Korea), in cooperation with the Republic of Chile. The Joint ALMA Observatory is operated by ESO, AUI/NRAO and NAOJ.


\begin{deluxetable*}{cccccccc}
\movetableright=0.1in
\tablecaption{Properties of Continuum Sources in HOPS 373SW\label{tb:gauss_cont}}
\tablehead{
\colhead{} & \colhead{R.A.}& \colhead{Dec.}& \colhead{b$_{maj}$}&\colhead{b$_{min}$} & \colhead{Flux density} & \colhead{Peak Intensity}& \colhead{Mass$^a$}\\\colhead{} & \colhead{}& \colhead{}& \colhead{mas}&\colhead{mas} & \colhead{mJy} & \colhead{mJy beam$^{-1}$}& \colhead{$\times 10^{-3}$ \msun}}
\startdata
 SW &  05:46:30.90483 & -00:02:35.19817 &  71.1 & 61.9   & 90.4 & 50.37  & 20.8  \\
    & (0$''$.00046)   & (0$''$.00048)   & ( 2.6) & (2.6) & (1.4)& (0.55)& (0.3) \\
 NE &  05:46:31.09929 & -00:02:33.0282  &  145.8 & 136.7 & 75.1 & 16.57 & 17.3 \\
    & (0$''$.00179)   & (0$''$.00175)   & (7.4)  & (7.8) & (2.3)& (0.43)& (0.5)\\
\hline
\multicolumn{8}{l}{\footnotesize The peak positions, deconvolved source sizes, continuum flux density, and peak intensity are measured with the task, }\\
\multicolumn{8}{l}{\footnotesize  imfit in CASA. }\\
\multicolumn{8}{l}{\footnotesize The value in parenthesis is the 1 $\sigma$ error.} \\
\multicolumn{8}{l}{\footnotesize $^a$The mass of each continuum source is calculated by the equation (1).}
\enddata
\end{deluxetable*}

\clearpage

\startlongtable
\begin{deluxetable*}{ccccccc}
\tabletypesize{\scriptsize}
\tablecaption{Identified Lines \label{tb:identified_lines}}
\tablehead{
\colhead{Line No.} &  \colhead{Formula}& \colhead{Name}&  \colhead{Frequency [GHz]}& \colhead{Transition} &  \colhead{Einstein-A [log$_{10}$A]} & \colhead{E$_u$ [K]}}
\startdata
\multicolumn{7}{c}{spw1}\\
\hline
1 & CD$_{3}$OH  & Methanol & 335.34538560 (1.10e-03) & 11(0,11)-10(1,9)E & -3.68529 & 133.34479 \\
2 & CH$_{3}$CHO  & Acetaldehyde & 335.3587225 (2.83e-05) & 18(0,18)-17(0,17)A,v$_{\rm t}$=0 & -2.87128 & 154.85102 \\
3 & CH$_{3}$CHO v$_{\rm t}$=1 & Acetaldehyde &	335.38246150 (3.91e-02) & 18(0,18)-17(0,17)E v$_{\rm t}$=1 & -2.86455 & 361.480 \\
4 & HDO  &	Water &	 335.3955 (2.6e-05) & 3(3,1)-4(2,2) & -4.58367 & 335.26718 \\
5 & C$_{2}$H$_{5}$OH  &	Ethanol & 335.4406247 (1.84e-05) & 33(20,13)-34(17,18),anti & -4.93008 & 877.18901  \\
  & C$_{2}$H$_{5}$OH  &	Ethanol & 335.4406247 (1.84e-05) & 33(20,14)-34(17,17),anti & -4.93008 & 877.18901  \\
6 & HCOCH$_{2}$OH   &  Glycolaldehyde & 335.4410822 (7.5e-06) & 34(14,20)-34(13,21) & -3.23609 & 450.0\\
  & HCOCH$_{2}$OH   &  Glycolaldehyde & 335.441595 (7.5e-06) & 34(14,21)-34(13,22) & -3.23609 & 450.00002\\
\hline
\multicolumn{7}{c}{spw2}\\
\hline
7 & CHD$_{2}$OH  &	Methanol  &	336.5034229 (2.6e-02) & 17(4)$^-$-17(3)$^+$ & -4.11594 & 356.38737 \\
8 & C$_{2}$H$_{5}$OH  &	Ethanol  &	336.5724107 (2e-06) & 19(3,16)-18(3,15),g- & -3.50357 & 232.33744 \\
9 & CH$_{3}$OH v$_{\rm t}$=2   &	Methanol & 336.60588900 (1.3e-02) & 7(1)$^+$-6(1)$^+$ v$_{\rm t}$=2 &	-3.78622 &	747.40430 \\
10 & C$_{2}$H$_{5}$OH  &	Ethanol  &	336.6264009 (3.6e-06) & 19(2,18)-18(1,17),anti & -3.56713 & 162.61385 \\
11 & CHD$_{2}$OH  &	Methanol  &	336.6270485 (3.0e-02) & 18(4)$^-$-18(3)$^+$ & -4.10425 & 392.42430 \\
12 & SO$_{2}$  &	Sulfur Dioxide & 336.6695774 (3.9e-06) & 16(7,9)-17(6,12) & -4.23374 & 245.11422 \\
\hline
\multicolumn{7}{c}{spw3}\\
\hline
13 & CD$_{3}$OH  & Methanol & 337.0183286 (1.3e-03) & 21(3,18)-21(2,19) & -3.69186 & 461.01808 \\
14 & CH$_{3}$OH v$_{\rm t}$=2   &	Methanol &	337.02191700 (4.1e-02)   & 7(-3)-6(-3)EE v$_{\rm t}$=2 & -3.86646  & 979.70052 \\ 
15 & CH$_{3}$OH v$_{\rm t}$=2   &	Methanol &	337.02957300 (1.8e-02) &	7(2)$^+$-6(2)$^+$ v$_{\rm t}$=2 & -4.23374 & 941.37640 \\
  & CH$_{3}$OH v$_{\rm t}$=2   &	Methanol &	337.02966200 (1.8e-02)  &	7(2)$^-$-6(2)$^-$ v$_{\rm t}$=2  & -4.23374 & 941.37641\\
16 & CH$_{2}$DOH &	Methanol &	337.0623909 (8.6e-06) & 8(2,7)-7(1,6),e0 & -5.02093 & 93.30945\\
17 & C$^{17}$O &	Carbon Monoxide &	337.0611298 (1e-05) & J=3-2 & -5.6344 & 32.35323 \\
18 & CH$_{3}$OH v$_{\rm t}$=2   &	Methanol &	337.06628100 (2.0e-02) & 7(4)-6(4)E v$_{\rm t}$=2 & -3.95078  & 950.98626 \\
19 & CH$_{3}$OH v$_{\rm t}$=2   &	Methanol &	337.07870700 (1.1e-02) & 7(5)-6(5)E v$_{\rm t}$=2 & -4.07598  & 909.35465 \\
20 & CHD$_{2}$OH  &	Methanol  &	337.0983031 (2.1e-02) & 4(4)$^+$-3(3)$^+$ & -3.99140 & 75.09492 \\
   & CHD$_{2}$OH  &	Methanol  &	337.0983227 (2.1e-02) & 4(4)$^-$-3(3)$^-$ & -3.99140 & 75.09492 \\
21 & CH$_{3}$OH v$_{\rm t}$=2   &	Methanol &	337.09891800 (1.8e-02) & 7(5)$^+$-6(5)$^+$ v$_{\rm t}$=2 & -4.08911  & 935.15562 \\
22 & CH$_{3}$OH v$_{\rm t}$=2   &	Methanol &	337.11386800 (1.1e-02) & 7(1)-6(1)E v$_{\rm t}$=2 & -3.78173  & 863.67831 \\
\hline
\multicolumn{7}{c}{spw4}\\
\hline
23 & CH$_{3}$OCHO &		Methyl Formate & 338.338184 (0.0002) & 27(8,19)-26(8,18)E & -3.26914 & 267.18358\\
24 & CH$_{3}$OH &	Methanol	& 338.344588 (5e-06) & 7(1,7)-6(1,6)E,v$_{\rm t}$=0 & -3.77809 & 70.55025 \\
25 & CH$_{3}$OCHO &		Methyl Formate	& 338.355792 (0.0001) & 27(8,19)-26(8,18)A & -3.26882 & 267.18572 \\
26 & CH$_{3}$OCHO &		Methyl Formate	& 338.396318 (0.0001) & 27(7,21)-26(7,20)E & -3.25943 & 257.74503 \\
27 & CH$_{3}$OH &	Methanol	& 338.40461 (5e-06) & 7(-6,2)-6(-6,1)E,v$_{\rm t}$=0 & -4.34503 & 243.78966 \\
28 & CH$_{2}$DOH &	Methanol	& 338.4071251 (1.25e-05) & 4(4,0)-3(3,0),e0 & -4.11124 & 84.40172 \\
   & CH$_{2}$DOH &	Methanol	& 338.4071284 (1.25e-05) & 4(4,1)-3(3,1),e0 & -4.11124 & 84.40172 \\
29 & CH$_{3}$OH &	Methanol	& 338.408698 (5e-06) & 7(0,7)-6(0,6)A,v$_{\rm t}$=0 & -3.76908 & 64.98055 \\
30 & C$_{2}$H$_{5}$OH  &	Ethanol	& 338.4116101 (3.8e-06) & 17(7,10)-17(6,11),anti & -3.65839 & 190.04683\\
31 & CH$_{3}$OCHO &		Methyl Formate	& 338.414116 (0.0001) & 27(7,21)-26(7,20)A & -3.25931 & 257.74703 \\
32 & C$_{2}$H$_{5}$OH  &	Ethanol	&  338.4168741 (3.8e-06) & 17(7,11)-17(6,12),anti & -3.65838 & 190.0468 \\
33 & CH$_{3}$OH &	Methanol	& 338.430975 (5e-06) & 7(6,1)-6(6,0)E,v$_{\rm t}$=0 & -4.34269 & 253.94589 \\
34 & CH$_{3}$OH &	Methanol	& 338.442367 (5e-06) & 7(6,2)-6(6,1)A,v$_{\rm t}$=0 & -4.3436 & 258.69481 \\
35 & CH$_{3}$OH &	Methanol	& 338.456536 (5e-06) & 7(5,3)-6(5,2)E,v$_{\rm t}$=0 & -4.07838 & 188.99765 \\
36 & CH$_{2}$DOH &	Methanol	& 338.4625377 (9e-06) & 9(1,8)-8(2,6),o1 & -4.23143 & 120.01391 \\
37 & CH$_{3}$OH &	Methanol	& 338.475226 (5e-06) & 7(-5,2)-6(-5,1)E,v$_{\rm t}$=0 & -4.0781 & 201.05874 \\
38 & CH$_{3}$OH &	Methanol	& 338.486322 (5e-06) & 7(5,2)-6(5,1)A,v$_{\rm t}$=0 & -4.07634 & 202.88349 \\
39 & CH$_{3}$OH &	Methanol	& 338.504065 (5e-06) & 7(4,4)-6(4,3)E,v$_{\rm t}$=0 & -3.94039 & 152.89245 \\
40 & CH$_{3}$OH &	Methanol	& 338.512632 (5e-06) & 7(4,4)-6(4,3)A,v$_{\rm t}$=0 & -3.93973 & 145.3317 \\
   & CH$_{3}$OH &	Methanol	& 338.512644 (5e-06) & 7(4,3)-6(4,2)A,v$_{\rm t}$=0 & -3.93973 & 145.3317 \\
   & CH$_{3}$OH &	Methanol	& 338.512853 (5e-06) & 7(2,6)-6(2,5)A,v$_{\rm t}$=0 & -3.80284 & 102.70215 \\
41 & CH$_{3}$OH &	Methanol	& 338.530257 (5e-06) & 7(-4,3)-6(-4,2)E,v$_{\rm t}$=0 & -3.93784 & 160.98995 \\
42 & CH$_{3}$OH &	Methanol	& 338.540826 (5e-06) & 7(3,5)-6(3,4)A,v$_{\rm t}$=0 & -3.8573 & 114.79245 \\
43 & CH$_{3}$OH &	Methanol	& 338.543152 (5e-06) & 7(3,4)-6(3,3)A,v$_{\rm t}$=0 & -3.8573 & 114.79257 \\
44 & CH$_{3}$OH &	Methanol	& 338.559963 (5e-06) & 7(3,4)-6(3,3)E,v$_{\rm t}$=0 & -3.85418 & 127.70603 \\
45 & CH$_{3}$OH &	Methanol	& 338.583216 (5e-06) & 7(-3,5)-6(-3,4)E,v$_{\rm t}$=0 & -3.85587 & 112.70842 \\
46 & SO$_{2}$  &	Sulfur Dioxide	& 338.6118078 (3.5e-06) & 20(1,19)-19(2,18) & -3.54241 & 198.8775 \\
47 & CH$_{3}$OH &	Methanol	& 338.614936 (5e-06) & 7(-1,6)-6(-1,5)E,v$_{\rm t}$=0 & -3.76662 & 86.05157 \\
48 & HCOCH$_{2}$OH  &	Glycolaldehyde	&  338.6285049 (6.4e-06) & 29(14,15)-29(13,16) & -3.26414 & 360.71888\\
   & HCOCH$_{2}$OH  &	Glycolaldehyde	&  338.6285126 (6.4e-06) & 29(14,16)-29(13,17) & -3.26414 & 360.71888\\
49 & CH$_{3}$OH &	Methanol	& 338.639802 (5e-06) & 7(2,5)-6(2,4)A,v$_{\rm t}$=0 & -3.80236 & 102.71572 \\
50 & CHD$_{2}$OH  &	Methanol    & 338.6594300 (1.52e-02) & 5(2)$^-$-5(1)$^+$ & -3.79102 & 49.84884 \\
51 & C$_{2}$H$_{5}$OH  &	Ethanol	& 338.6717299 (3.9e-06) & 16(7,9)-16(6,10),anti & -3.66791 & 175.74598\\
52 & C$_{2}$H$_{5}$OH  &	Ethanol	& 338.6742587 (3.9e-06) & 16(7,10)-16(6,11),anti & -3.66791 & 175.74596\\
53 & CH$_{3}$OH &	Methanol	& 338.721693 (5e-06) & 7(-2,6)-6(-2,5)E,v$_{\rm t}$=0 & -3.80944 & 87.25749 \\
   & CH$_{3}$OH &	Methanol	& 338.722898 (5e-06) & 7(2,5)-6(2,4)E,v$_{\rm t}$=0 & -3.80434 & 90.91288 \\
54 & CH$_{2}$DOH &	Methanol	& 338.7402165 (1.67e-05) & 23(1,22)-23(0,23),e0 & -3.92503 & 600.09684 \\
55 & $^{13}$CH$_{3}$OH &	Methanol	& 338.759948 (5e-05) & 13(0,13)-12(1,12)++ & -3.66185 & 205.94761 \\
\hline
\multicolumn{7}{c}{spw5}\\
\hline
56 & CH$_{3}$CHO  & Acetaldehyde &		346.9575559 (2.84e-05) & 18(7,12)-17(7,11)A,v$_{\rm t}$=0 & -2.89625 & 268.60294 \\
   & CH$_{3}$CHO  & Acetaldehyde &		346.9575577 (2.84e-05) & 18(7,11)-17(7,10)A,v$_{\rm t}$=0 & -2.89625 & 268.60294 \\
57 & C$_{2}$H$_{5}$OH  &	Ethanol	 & 346.9626034 (3.2e-06) & 21(0,21)-20(1,20),anti & -3.35133 & 185.84605	\\
58 & C$_{2}$H$_{5}$OH  &	Ethanol & 346.9745486 (6.7e-06) & 17(3,15)-16(2,15),g+ & -4.00072 & 195.1701 \\
59 & CH$_{3}$CHO  & Acetaldehyde	 &	 	346.9955324 (2.82e-05) & 18(7,12)-17(7,11)E,v$_{\rm t}$=0 & -2.89625 & 268.5688	 \\
60 & CH$_{3}$CHO  & Acetaldehyde &		347.0715471 (2.58e-05) & 18(6,13)-17(6,12)A,v$_{\rm t}$=0 & -2.87569 & 239.3968  \\
   & CH$_{3}$CHO  & Acetaldehyde &		347.0716844 (2.58e-05) & 18(6,12)-17(6,11)A,v$_{\rm t}$=0 & -2.87569 & 239.39681  \\
61 & CH$_{3}$CHO  & Acetaldehyde	 &		347.0904015 (2.57e-05) & 18(6,12)-17(6,11)E,v$_{\rm t}$=0 & -2.87577 & 239.39469 \\	
62 & CH$_{3}$CHO  & Acetaldehyde	 &	 	347.1326859 (2.58e-05) & 18(6,13)-17(6,12)E,v$_{\rm t}$=0 & -2.87563 & 239.3183 \\
63 & C$_{2}$H$_{5}$OH  &	Ethanol & 347.1472094 (1.9e-06) & 20(6,15)-19(6,14),g+ & -3.45039 & 275.45993	 \\
64 & C$_{2}$H$_{5}$OH  &	Ethanol & 347.1579974 (1.9e-06) & 20(6,14)-19(6,13),g+ & -3.45038 & 275.46117	 \\
65 & CH$_{3}$CHO  & Acetaldehyde	 & 347.1695196 (3.96e-05) & 19(0,19)-18(1,18)A,v$_{\rm t}$=0 & -3.66327 & 171.81276 \\
66 & CH$_{3}$CHO v$_{\rm t}$=1 & Acetaldehyde &	347.18241320 (2.99e-02) & 18(4,14)-17(4,13)E v$_{\rm t}$=1 & -2.84527 & 400.374 \\
67 & CH$_{2}$DOH &	Methanol &		347.1880746 (9.2e-06) & 8(5,3)-9(4,6),e1 & -5.06349 & 185.14136 \\	
   & CH$_{2}$DOH &	Methanol &		347.1881494 (9.2e-06) & 8(5,4)-9(4,5),e1 & -5.06349 & 185.14137	 \\
68 & $^{13}$CH$_{3}$OH &		Methanol &	347.188283 (6.4e-05) & 14(1,13)-14(0,14)-+ & -3.36087 & 254.25506 \\	
69 & CH$_{3}$CHO v$_{\rm t}$=1	& Acetaldehyde & 347.21679780 (2.88e-02) & 18(5,13)-17(5,12)E v$_{\rm t}$=1 & -2.85918 & 420.436\\
70 & CH$_{2}$DOH &	Methanol &		347.2229918 (9.7e-06) & 20(4,16)-20(3,18),e1 & -3.87846 & 521.63257	 \\
71 & CH$_{3}$CHO  & Acetaldehyde &	347.2403963 (3.95e-05) & 19(0,19)-18(1,18)E,v$_{\rm t}$=0 & -3.66518 & 171.88508 \\	
72 & CH$_{3}$CHO v$_{\rm t}$=1	& Acetaldehyde & 347.25182200 (3.08e-02) & 18(5,14)-17(5,13)E v$_{\rm t}$=1 & -2.85886 & 419.668	\\
73 & CH$_{3}$CHO  & Acetaldehyde &		347.288264 (2.48e-05) & 18(5,14)-17(5,13)A,v$_{\rm t}$=0 & -2.85868 & 214.69491 \\	
74 & CH$_{3}$CHO  & Acetaldehyde &		347.2948735 (2.48e-05) & 18(5,13)-17(5,12)A,v$_{\rm t}$=0 & -2.85857 & 214.69566 \\	
75 & SiO  & Silicon Monoxide & 347.330631 (3.475e-04) & J=8-7 & -2.65578 & 75.01697 \\
76 & CH$_{3}$CHO  & Acetaldehyde &		347.3457104 (2.48e-05) & 18(5,13)-17(5,12)E,v$_{\rm t}$=0 & -2.8586 & 214.63811	 \\
77 & CH$_{3}$CHO  & Acetaldehyde &		347.3492783 (2.48e-05) & 18(5,14)-17(5,13)E,v$_{\rm t}$=0 & -2.85853 & 214.60878	 \\
78 & C$_{2}$H$_{5}$OH  &	Ethanol & 347.3510886 (6.2e-06) & 14(3,12)-13(2,11),anti & -3.80615 & 99.66183 \\
79 & CH$_{2}$DOH &	Methanol &		347.3703636 (1.32e-05) & 22(2,20)-22(2,21),e1 & -7.29364 & 571.62552	 \\
   & CH$_{2}$DOH &	Methanol &		347.3711668 (9.4e-06) & 19(4,16)-19(3,16),e1 & -3.88761 & 478.84197	 \\
\hline
\multicolumn{7}{c}{spw6}\\
\hline
80 & CH$_{2}$DOH &	Methanol &		349.1838267 (8.2e-06) & 13(4,9)-13(3,11),e1 & -3.91044 & 266.96107 \\
81 & CH$_{3}$CN	 &	Methyl Cyanide &		349.2123106 (2e-07) & 19(6)-18(6) & -2.63745 & 424.70492 \\
82 & CH$_{2}$DOH &	Methanol &		349.2422265 (1.38e-05) & 21(2,20)-21(1,21),o1 & -3.85689 & 529.71963\\
83 & C$_{2}$H$_{5}$OH  & Ethanol & 349.2844104 (5.5e-06) & 9(2,8)-8(1,8),g- & -4.06388 & 103.72509\\	
84 & CH$_{3}$CN	 &	Methyl Cyanide &		349.2860057 (2e-07) & 19(5)-18(5) & -2.62274 & 346.22396 \\
85 & CH$_{3}$CHO v$_{\rm t}$=1 & Acetaldehyde & 349.32035150 (3.33e-02) & 18(1,17)-17(1,16)E v$_{\rm t}$=1 & -2.81559 & 369.252\\	
86 & CD$_{3}$OH  &  Methanol & 349.3315497 (7.0e-04) & 11(0,11)-10(1,9)E & -3.72285 & 94.23705 \\
87 & CH$_{2}$DOH &	Methanol  &	349.333979 (8.2e-06) & 12(4,9)-12(3,9),e1 & -3.91862 & 239.13136 \\	
88 & CH$_{3}$CN	 &	Methyl Cyanide &	349.3463428 (2e-07) & 19(4)-18(4) & -2.61106 & 281.98778 \\	
89 & CH$_{2}$DOH &	Methanol &	349.3561319 (8.1e-06) & 12(4,8)-12(3,10),e1 & -3.9182 & 239.13127 \\	
90 & CH$_{3}$CN	 &	Methyl Cyanide &	349.3932971 (2e-07) & 19(3)-18(3) & -2.60218 & 232.01121 \\	
\hline
\multicolumn{7}{c}{spw7}\\
\hline
91 & CH$_{2}$DOH &	Methanol &		350.0902424 (9.4e-06) & 5(4,2)-5(3,2),e1 & -4.12044 & 104.24014 \\
   & CH$_{2}$DOH &	Methanol &		350.0903841 (9.4e-06) & 5(4,1)-5(3,3),e1 & -4.12044 & 104.24015 \\
92 & $^{13}$CH$_{3}$OH &		Methanol &		350.103118 (5e-05) & 1(1,1)-0(0,0)++ & -3.48252 & 16.80241 \\
93 & CH$_{3}$CHO  & Acetaldehyde &	350.1334296 (2.75e-05) & 18(3,15)-17(3,14)A,v$_{\rm t}$=0 & -2.82551 & 179.20697 \\
   & CH$_{3}$CHO  & Acetaldehyde &	350.1343816 (2.75e-05) & 18(3,15)-17(3,14)E,v$_{\rm t}$=0 & -2.82596 & 179.17508 \\
94 & CH$_{2}$DOH &	Methanol &		350.1413021 (1.1e-05) & 4(4,1)-4(3,1),e1 & -4.29793 & 93.53285 \\
   & CH$_{2}$DOH &	Methanol &		350.1413378 (1.1e-05) & 4(4,0)-4(3,2),e1 & -4.29793 & 93.53285 \\
\hline
\multicolumn{7}{c}{spw8}\\
\hline
95 & CH$_{3}$CHO  & Acetaldehyde &	350.3628435 (2.73e-05) & 18(1,17)-17(1,16)E,v$_{\rm t}$=0 & -2.81493 & 163.45886	 \\
96 & C$_{2}$H$_{5}$OH  &	Ethanol & 350.3650648 (2e-06) & 20(4,16)-19(4,15),g+ & -3.41113 & 251.74648	 \\
97 & $^{13}$CH$_{3}$OH &		Methanol &		350.421585 (5e-05) & 8(1,7)-7(2,5) & -4.15313 & 102.61684	 \\
98 & CH$_{3}$OCHO &		Methyl Formate & 350.44225 (0.0001) & 28(8,21)-27(8,20)E & -3.21991 & 283.90675	 \\
99 & CH$_{3}$CHO  & Acetaldehyde & 	350.4457777 (2.74e-05) & 18(1,17)-17(1,16)A,v$_{\rm t}$=0 & -2.81489 & 163.41752	 \\
100 & CH$_{2}$DOH &	Methanol &		350.4538682 (7.7e-06) & 6(2,5)-5(1,5),e1 & -3.85626 & 71.55787	 \\
101 & CH$_{3}$OCHO &	Methyl Formate & 350.45758 (5e-05) & 28(8,21)-27(8,20)A & -3.21978 & 283.9095	 \\
102 & CHD$_{2}$OH  &	Methanol    & 350.4673839 (2.8e-02) & 16(1)$^+$-15(2)$^+$ & -4.20845 & 276.34756 \\
\hline
\multicolumn{7}{l}{Note. Species with the same numbering are observed as a blended line.}
\enddata
\end{deluxetable*}

\begin{deluxetable*}{ccccccc}
\tablecaption{Tentatively Identified Lines} \label{tb:tentative_detection}
\tablehead{
\colhead{Line No.} &  \colhead{Formula}& \colhead{Name}&  \colhead{Frequency [GHz]}& \colhead{Transition} &  \colhead{Einstein-A [log$_{10}$A]} & \colhead{E$_u$ [K]}}
\startdata
\multicolumn{7}{c}{spw1}\\
\hline
1 & CH$_{2}$DCHO & Acetaldehyde & 335.40589500 (1.0e-02) & 19(0,19)-18(0,18) & -2.87877 & 163.435 \\
\hline
\multicolumn{7}{c}{spw2}\\
\hline
2 & HCCCN & Cyanoacetylene & 336.52008400 (1.0e-02) & J=37-36 & -2.51613 & 306.906\\
\hline
\multicolumn{7}{c}{spw3}\\
\hline
3 & CH$_{3}$CHO v$_{\rm t}$=2 & Acetaldehyde & 337.08157220 (1.43e-01)  & 18(1,18)-17(1,17) & -2.88207 & 526.134\\
\hline
\multicolumn{7}{c}{spw4}\\
\hline
4 & CH$_{3}$CHO v$_{\rm t}$=2 & Acetaldehyde & 338.31785810 (1.61e-01) & 18(5,14)-17(5,13) & -2.86934 & 605.150\\
5 & $^{34}$SO$_{2}$ & Sulfur Dioxide & 338.32035640 (4.2e-03) & 13(2,12)-12(1,11) & -3.64397 & 92.449\\
\hline
\multicolumn{7}{c}{spw5}\\
\hline
6 & CH$_{3}$OD & Methanol & 347.308432 (1.39e-04) & 5(4)$^-$-6(3)$^-$ & --\tablenotemark{a} & 104.311\\
\hline
\multicolumn{7}{c}{spw8}\\
\hline
7 & CD$_{3}$OH v$_{\rm t}$=1 & Methanol & 350.430949 (1.0e-03) & 9(1,8)A-8(1,7)A & -3.71041 & 384.41981\\
\enddata
\tablenotetext{a}{The information the CH$_{3}$OD line is from \citet{Anderson1988ApJS...67..135A} and \citet{Duan2003JMoSp.218...95D}, which do not provide its Einstein-A coefficient.}
\end{deluxetable*}

\begin{deluxetable*}{cccccc}
\tablecaption{Results of XCLASS fittings} \label{tb:xclass_results}
\tablehead{
\colhead{Species} &  \colhead{Formula}& \colhead{Column density\tablenotemark{a} [cm$^{-2}$]}&  \colhead{T$_{rot}$ [K]}& \colhead{X({\it w.r.t.} H$_2$)} & \colhead{X({\it w.r.t.} CH$_3$OH)\tablenotemark{b}}}
\startdata
\multicolumn{6}{c}{Identified}\\
\hline
Methanol & CH$_{3}$OH &  8.04$_{-0.69}^{+1.06}\times$10$^{17}$ & 75.9$_{-4.24}^{+0.53}$ & 9.03$\times$10$^{-8}$ &  --\\
         & CH$_{3}$OH v$_{\rm t}$=2 &  1.11$_{-0.16}^{+2.97}\times$10$^{18}$ & 200$_{-11.3}^{+6.04}$ & 1.25$\times$10$^{-7}$ &  3.61$\times$10$^{-1}$\\
         & CD$_{3}$OH &  6.20$_{-1.84}^{+1.65}\times$10$^{16}$ & 126$_{-2.07}^{+2.53}$ & 6.97$\times$10$^{-9}$ &  2.02$\times$10$^{-2}$\\
         & $^{13}$CH$_{3}$OH & 5.11$_{-1.33}^{+0.70}\times$10$^{16}$ & 98.0$_{-4.19}^{+2.79}$ & 5.75$\times$10$^{-9}$ &  1.67$\times$10$^{-2}$\\
         & CH$_{2}$DOH & 4.03$_{-0.59}^{+0.72}\times$10$^{17}$ & 199$_{-5.85}^{+4.79}$ & 4.53$\times$10$^{-8}$ & 1.31$\times$10$^{-1}$\\
         & CHD$_{2}$OH & 3.03$_{-0.42}^{+0.98}\times$10$^{17}$ & 237$_{-4.78}^{+4.78}$ & 3.40$\times$10$^{-8}$ & 9.87$\times$10$^{-2}$\\
Acetaldehyde & CH$_{3}$CHO & 1.97$_{-0.54}^{+0.37}\times$10$^{16}$ & 129$_{-10.4}^{+4.47}$ & 2.21$\times$10$^{-9}$ & 6.41$\times$10$^{-3}$\\
    &    CH$_{3}$CHO v$_{\rm t}$=1 & 2.06$_{-0.56}^{+0.60}\times$10$^{16}$ & 210$_{-47.1}^{+47.1}$ & 2.31$\times$10$^{-9}$ & 6.72$\times$10$^{-3}$\\
Methyl Formate & CH$_{3}$OCHO & 1.53$_{-0.22}^{+0.41}\times$10$^{17}$ & 445$_{-6.43}^{+3.46}$ & 1.71$\times$10$^{-8}$ & 4.98$\times$10$^{-2}$\\
Ethanol & C$_{2}$H$_{5}$OH & 5.07$_{-0.68}^{+1.54}\times$10$^{16}$ & 116$_{-4.56}^{+6.84}$ & 5.70$\times$10$^{-9}$ & 1.65$\times$10$^{-2}$\\
Methyl Cyanide & CH$_{3}$CN & 2.24$_{-0.42}^{+0.52}\times$10$^{15}$ & 207$_{-7.35}^{+10.8}$ & 2.51$\times$10$^{-10}$ & 7.29$\times$10$^{-4}$\\
Glycolaldehyde & HCOCH$_{2}$OH & 2.09$_{-0.53}^{+0.35}\times$10$^{16}$ & 116$_{-2.90}^{+6.78}$ & 2.35$\times$10$^{-9}$ & 6.81$\times$10$^{-3}$\\
Water & HDO\tablenotemark{c} &  1.40$_{-0.12}^{+0.17}\times$10$^{17}$ & 100 & 1.58$\times$10$^{-8}$ & 4.56$\times$10$^{-2}$\\
Sulfur Dioxide & SO$_{2}$ & 2.92$_{-0.68}^{+0.45}\times$10$^{17}$ & 62.0$_{-1.04}^{+3.65}$ & 3.29$\times$10$^{-8}$ & 9.51$\times$10$^{-2}$\\
Silicon Monoxide & SiO\tablenotemark{c} & 4.96$_{-0.24}^{+0.60}\times$10$^{13}$ & 100 & 5.58$\times$10$^{-12}$ & 1.62$\times$10$^{-5}$\\
\hline
\multicolumn{6}{c}{Tentatively Identified}\\
\hline
Acetaldehyde & CH$_{2}$DCHO\tablenotemark{c} & 6.24$_{-0.24}^{+0.48}\times$10$^{15}$ & 100 & 7.01$\times$10$^{-10}$ & 2.03$\times$10$^{-3}$\\
Cyanoacetylene & HCCCN\tablenotemark{c} & 1.06$_{-0.22}^{+0.06}\times$10$^{15}$ & 100 & 1.19$\times$10$^{-10}$ & 3.45$\times$10$^{-4}$\\
Sulfur Dioxide & $^{34}$SO$_{2}$\tablenotemark{c} & 4.69$_{-0.66}^{+0.29}\times$10$^{15}$ & 100 & 5.27$\times$10$^{-10}$ & 1.53$\times$10$^{-3}$\\
Methanol & CD$_{3}$OH v$_{\rm t}$=1\tablenotemark{c} &  3.73$_{-0.34}^{+0.58}\times$10$^{17}$ & 100 & 4.19$\times$10$^{-8}$ & 1.21$\times$10$^{-1}$\\
\hline
\multicolumn{6}{l}{Note. Linewidths were fixed as 2 km s$^{-1}$.}
\enddata
\tablenotetext{a}{The uncertainty of column density corresponds to the 1 $\sigma$ error.}
\tablenotetext{b}{X($w.r.t$ CH$_3$OH) was calculated using the derived column density of $^{13}$CH$_3$OH and the $^{12}$C/$^{13}$C ratio of 60.}
\tablenotetext{c}{Only one transition was detected for these molecules, so the temperature was fixed as 100 K in the XCLASS fitting.}
\end{deluxetable*}

\begin{deluxetable*}{lcc}[]
\tablecaption{D/H ratio}
\label{table:deut}
\tablehead{\colhead{} & \colhead{Observed Ratio} &  \colhead{D/H [$\%$]
\tablenotemark{a}
}}
\startdata
    CH$_{2}$DOH/CH$_{3}$OH\tablenotemark{b}
    & 0.36 & 12 \\
    CHD$_{2}$OH/CH$_{3}$OH & 0.27 & 30 \\ 
    CD$_{3}$OH/CH$_{3}$OH & 0.06 & 38 \\
    CH$_{2}$DCHO/CH$_{3}$CHO & 0.32 & 11 \\
    \enddata
\tablenotetext{a}{Statistical correction has applied to measure the D/H ratio (\citealt{Parise04}; Appendix B of \citealt{Manigand2019})}
\tablenotetext{b}{The CH$_3$OH column density was calculated using the CH$_3$OH v$_{t}$=2 lines.}
\end{deluxetable*}


\begin{deluxetable*}{ccccccccc}
\tablecaption{X($w.r.t$ CH$_3$OH) in Various Sources \label{tb:comparison}}
\tabletypesize{\scriptsize}
\tablehead{\colhead{Molecules} &  \multicolumn{8}{c}{Sources}}
\startdata
{} &
{HOPS 373SW\tablenotemark{a}}& {ALMASOP\tablenotemark{a,b}}& {HH 212\tablenotemark{a,c}}& {V883 Ori\tablenotemark{a,d}} & {IRAS16293B\tablenotemark{e}}& {PEACHES\tablenotemark{f}}& {BHR71\tablenotemark{g}}& {L483\tablenotemark{a,h}}\\
\hline
CH$_{2}$DOH &
1.3$_{-0.4}^{+0.3}\times$10$^{-1}$ & 
2.9$\times$10$^{-2}$ & 
1.0$_{-0.2}^{+0.2}\times$10$^{-1}$ & 
2.0$_{-0.1}^{+0.1}\times$10$^{-2}$ & 
7.1$_{-2.1}^{+2.1}\times$10$^{-2}$ & 
-- & 
2.9$\times$10$^{-2}$ & 
2.7$_{-0.8}^{+0.8}\times$10$^{-2}$  
\\
CH$_{3}$CHO &
6.4$_{-2.4}^{+1.5}\times$10$^{-3}$ & 
1.7$\times$10$^{-3}$ & 
9.6$_{-1.3}^{+1.3}\times$10$^{-3}$ & 
2.0$_{-0.1}^{+0.1}\times$10$^{-2}$ & 
1.2$_{-0.4}^{+0.4}\times$10$^{-2}$ & 
8.9$_{-6.7}^{+27.6}\times$10$^{-2}$ & 
-- & 
5.3$_{-1.5}^{+1.5}\times$10$^{-3}$  
\\
CH$_{3}$OCHO & 
5.0$_{-1.5}^{+1.5}\times$10$^{-2}$ & 
1.6$\times$10$^{-2}$ & 
2.1$_{-0.5}^{+0.5}\times$10$^{-2}$ & 
7.5$_{-0.1}^{+0.1}\times$10$^{-2}$ & 
2.6$_{-0.8}^{+0.8}\times$10$^{-2}$ & 
2.2$_{-1.5}^{+4.4}\times$10$^{-1}$ & 
1.8$\times$10$^{-2}$ & 
8.7$_{-2.5}^{+2.5}\times$10$^{-3}$  
\\ 
C$_{2}$H$_{5}$OH & 
1.7$_{-0.5}^{+0.6}\times$10$^{-2}$ & 
2.4$\times$10$^{-2}$ & 
1.7$_{-0.3}^{+0.3}\times$10$^{-2}$ & 
-- & 
2.3$_{-0.7}^{+0.7}\times$10$^{-2}$ & 
1.7$_{-1.2}^{+3.5}\times$10$^{-1}$ & 
5.4$\times$10$^{-3}$ & 
6.7$_{-1.9}^{+1.9}\times$10$^{-3}$ 
\\ 
CH$_{3}$CN & 
7.3$_{-2.3}^{+2.0}\times$10$^{-4}$ & 
-- & 
-- & 
7.8$_{-0.1}^{+0.1}\times$10$^{-4}$ & 
4.0$_{-1.2}^{+1.0}\times$10$^{-3}$ & 
1.2$_{-0.8}^{+2.6}\times$10$^{-2}$ & 
-- & 
--  
\\ 
HCOCH$_{2}$OH & 
6.8$_{-2.5}^{+1.5}\times$10$^{-3}$ & 
-- & 
-- & 
-- & 
3.4$_{-1.0}^{+1.0}\times$10$^{-3}$ & 
1.4$_{-0.9}^{+2.8}\times$10$^{-2}$ & 
-- & 
--  
\\
HDO & 
4.6$_{-1.2}^{+0.8}\times$10$^{-2}$ & 
--& 
--& 
1.2$\times$10$^{-2}$ & 
--& 
--&
--& 
-- \\ 
\hline\hline
{Molecules} & \colhead{HOPS 373SW\tablenotemark{a}} & \multicolumn{7}{c}{5 sources in ALMASOP survey \tablenotemark{a}}\\
\hline
{} & {}& {G192}& {G205S1A}& {G208N1}& {G210WA}& {G211S} & {} & {} \\
{} & {} & {} & {HOPS-358-A}& {HOPS-87}& {HOPS-168-A}& {HOPS-288-A-A} & {} & {} \\
\hline
CH$_{2}$DOH & 
1.3$_{-0.4}^{+0.3}\times$10$^{-1}$ & 
2.5$_{-0.1}^{+0.1}\times$10$^{-2}$ & 
2.1$_{-0.7}^{+0.3}\times$10$^{-2}$ & 
3.0$_{-2.1}^{+1.2}\times$10$^{-2}$ & 
3.2$_{-0.2}^{+0.3}\times$10$^{-2}$ & 
4.4$_{-0.4}^{+0.1}\times$10$^{-2}$ & 
{} &
{}
\\
CH$_{3}$CHO &
6.4$_{-2.4}^{+1.5}\times$10$^{-3}$ & 
8.2$_{-0.3}^{+0.6}\times$10$^{-4}$ & 
1.2$_{-0.2}^{+0.1}\times$10$^{-3}$ & 
1.4$_{-0.1}^{+0.1}\times$10$^{-3}$ & 
1.4$_{-0.1}^{+0.1}\times$10$^{-3}$ & 
6.1$_{-0.8}^{+0.2}\times$10$^{-3}$ & 
{} &
{}
\\
CH$_{3}$OCHO & 
5.0$_{-1.5}^{+1.5}\times$10$^{-2}$ & 
1.3$_{-0.2}^{+0.1}\times$10$^{-2}$ & 
1.4$_{-0.4}^{+0.1}\times$10$^{-2}$ & 
1.2$_{-0.3}^{+0.1}\times$10$^{-2}$ & 
7.8$_{-0.8}^{+0.4}\times$10$^{-3}$ & 
2.1$_{-0.2}^{+0.1}\times$10$^{-2}$ & 
{} &
{}
\\ 
C$_{2}$H$_{5}$OH & 
1.7$_{-0.5}^{+0.6}\times$10$^{-2}$ & 
-- & 
-- & 
2.5$_{-0.1}^{+0.1}\times$10$^{-2}$ & 
-- & 
2.4$_{-0.1}^{+0.1}\times$10$^{-2}$ & 
{} &
{}
\\ 
CH$_{3}$CN\tablenotemark{i} & 
7.3$_{-2.3}^{+2.0}\times$10$^{-4}$ & 
-- & 
5.1$_{-0.2}^{+0.6}\times$10$^{-3}$ & 
6.8$_{-0.2}^{+0.1}\times$10$^{-3}$ & 
-- & 
9.6$_{-3.1}^{+0.4}\times$10$^{-3}$ & 
{} &
{}
\\
HCOCH$_{2}$OH & 
6.8$_{-2.5}^{+1.5}\times$10$^{-3}$ & 
-- & 
-- & 
3.6$_{-0.2}^{+0.2}\times$10$^{-3}$ & 
-- & 
-- & 
{} &
{}
\enddata
\tablenotetext{a}{X($w.r.t$ CH$_3$OH) was calculated using the derived column density of $^{13}$CH$_3$OH and the $^{12}$C/$^{13}$C ratio of 60.}
\tablenotetext{b}{X($w.r.t$ CH$_3$OH) are adopted from Figure 7 of \citet{Hsu22} and scaled down by a factor of 1.2 to adopt the $^{12}$C/$^{13}$C ratio of 60.}
\tablenotetext{c}{\citet{cflee19}.}
\tablenotetext{d}{\citet{jelee19}. The HDO abundance is derived from a tentative detection.}
\tablenotetext{e}{X($w.r.t$ CH$_3$OH) was calculated using the derived column density of CH$_3^{18}$OH and the $^{16}$O/$^{18}$O ratio of 560 \citep{Jorgensen18,Jorgensen20}. CH$_2$DOH abundance is adopted from \citep{Jorgensen18}.}
\tablenotetext{f}{\citet{Yang21}.}
\tablenotetext{g}{\citet{Yang20}. The abundances are estimated from the obtained column densities assuming T$_{ex}$=100 K.}
\tablenotetext{h}{\citet{Jacobsen19}.}
\tablenotetext{i}{X($w.r.t$ CH$_3$CN) of ALMASOP sources were calculated using the derived column density of $^{13}$CH$_3$CN and the $^{12}$C/$^{13}$C ratio of 60.}
\end{deluxetable*}

\clearpage

\bibliographystyle{aasjournal}
\bibliography{ms}{}


\end{document}